\documentclass[fleqn,usenatbib,useAMS]{mnras}

\usepackage{graphicx}	
\usepackage{amsmath}	
\usepackage{amssymb}	
\usepackage{multicol}        
\usepackage{bm}		
\usepackage{pdflscape}	

\usepackage{url,times,graphicx,hyperref,amsmath,amsfonts,amssymb,aas_macros,color,epsfig,epstopdf}
\usepackage{booktabs}
\usepackage[utf8]{inputenc}
\usepackage{ae,aecompl}
\usepackage{subcaption}
\usepackage{xcolor}
\usepackage{lipsum}

\defcitealias{Santos-Santos2025}{SS25}

\usepackage[T1]{fontenc}
\usepackage{ae,aecompl}

\usepackage{newtxtext,newtxmath}

\title[The missing ultra-faint satellites of the Milky Way]{The missing ultra-faint satellites of the Milky Way}

\author[Santos-Santos et al.]{
Isabel M.E.  Santos-Santos,$^{1}$\thanks{E-mail: isantossantos@aip.de}
Carlos S.  Frenk,$^{1}$
Julio F. Navarro,$^{2,3}$
\\
$^{1}$Institute for Computational Cosmology, Department of Physics, Durham University, South Road, Durham, DH1 3LE, UK\\
$^{2}$Department of Physics and Astronomy, University of Victoria, BC V8P 5C2, Canada\\
$^{3}$KU-KIST Green School, Korea University\\
}

\date{Accepted XXX. Received YYY; in original form ZZZ}

\pubyear{2025}

\begin{document}
\label{firstpage}
\pagerange{\pageref{firstpage}--\pageref{lastpage}}
\maketitle

\begin{abstract}
We combine the highest resolution N-body simulation of a $\sim10^{12}M_\odot$ $\Lambda$CDM halo (Aquarius-A) with the semi-analytic model GALFORM to study the full satellite population of Milky Way (MW)-like galaxies. The model assumes that galaxies form only in subhalos whose peak circular velocity exceeds the hydrogen-cooling threshold, and predicts that the total number of satellites ever accreted into the main halo, down to arbitrarily low luminosity, should not exceed a few hundred. The model tracks satellites even after their halos become unresolved ("orphan" galaxies), and includes a novel treatment of 
dark matter and stellar
tidal stripping which takes into account that $\Lambda$CDM subhalos  survive tidal disruption because of their cuspy density profiles. Our results match the massive end of the observed MW satellite mass function and predict that many ultra-faint dwarfs (UFDs) with $M_V>-5$ are missing from the current MW satellite census. The missing UFDs are predicted to avoid the innermost regions of the host, and have properties overlapping those of the recently-discovered ultra-faint compact MW satellites (UFCSs). If the faintest dwarfs are initially as compact as UFCSs, our results suggest that many UFCSs should be dark matter-dominated dwarfs with velocity dispersions $1-3$ km/s which survived disruption because they inhabit the  cusp of $\Lambda$CDM subhalos. UFCSs have predicted densities of  $10^{10}$-$10^{11}\,M_\odot/$kpc$^3$, higher than those of more extended UFDs. If confirmed, our results  provide strong support for the cuspy nature of $\Lambda$CDM halos and for the hydrogen-cooling threshold for  galaxy formation.
\end{abstract}


\section{Introduction}\label{sec:intro}
The $\Lambda$ Cold Dark Matter ($\Lambda$CDM) model describes a Universe in which dark matter clusters under gravity  to form equilibrium non-linear structures (``halos'') spanning a large range of masses \citep[from $\sim10^{-6}$ to $10^{15}$ M$_\odot$;][]{Wang2020}. In this framework, halos are the hosts of all galaxies, which form when gas bound to the halo  is able to cool and condense at the centre of these deep potential wells, allowing star formation to take place \citep{WhiteRees1978,WhiteFrenk1991}.

A robust prediction of $\Lambda$CDM is that such halos should exhibit a centrally steep  density profile whose shape is  independent of halo mass or cosmological parameters.  Cold dark matter halo profiles are well described by a simple formula with two parameters, halo mass and concentration, where the latter reflects the characteristic density of the halo, imprinted at the epoch of its assembly \citep[][hereafter ``NFW'']{Navarro1996,Navarro1997}.  

Dark matter halos grow with time via the hierarchical accretion of smaller structures \citep{FrenkWhite2012}. Traces of this accretion history survive to the present day as ``substructure'' made up of the remnants of subhalos accreted into a more massive host. Subhalos are expected to be the hosts of satellite galaxies, which means that, in principle, it should be possible to turn observations of satellite populations into useful probes of the clustering of dark matter on small scales and therefore a test of the $\Lambda$CDM model itself.

Because of its proximity, the satellite population of the Milky Way (MW) provides the best testbed for these theories. 
In the early 2000s, only the $\sim 11$ brightest ($M_V < -8$)  MW satellite galaxies were known. This number stood in stark contrast to the thousands of subhalos predicted to lie within the virial radius of a MW-mass halo in $\Lambda$CDM, giving rise to the well-known ``missing satellites'' problem \citep{Klypin1999,Moore1999}.

This problem can be readily resolved if luminous satellites form only in subhalos exceeding a certain mass threshold, where gas is able to cool radiatively and collapse to the centre to form a galaxy. Before the epoch of reionization, atomic gas can cool efficiently mainly in halos whose virial temperatures are high enough that atomic collisions can excite the hydrogen Ly-$\alpha$ transition. After reionization, the ultraviolet background heats the intergalactic medium to temperatures of order $\sim 2\times 10^4$ K, restricting cooling to halos with virial temperatures above this threshold 
\citep[e.g.][]{Efstathiou1992,Thoul1996,Tegmark1997,Gnedin2000,Okamoto2009,
Benitez-Llambay2020}. 
As a result, only halos above a certain redshift-dependent ``critical mass'' can host luminous galaxies and many low-mass halos and subhalos remain dark. Although the precise value of the critical mass and its redshift dependence are still debated, this overall scenario is widely accepted at present.

On the observational side, the advent of digital sky surveys has led to the discovery of  new MW satellites, bringing the number of confirmed satellite galaxies to $\sim 65$ today.  Over the last decade, in particular, the detection rate of ultra-faint stellar systems ($M_*<10^5$ M$_\odot$) has accelerated dramatically thanks to surveys such as the Sloan Digital Sky Survey \citep[SDSS][]{Abazajian2009,Willman2005,Belokurov2010}, Pan-STARRS1 \citep{Chambers2016}, the Dark Energy Survey \citep[DES; e.g.][]{Abbott2018,Bechtol2015,Drlica-Wagner2015}, the Hyper Suprime-Cam Subaru Strategic Program \citep[HSC-SSP;][]{Aihara2018}, DELVE \citep{Drlica-Wagner2021}, Gaia \citep[e.g.][]{Torrealba2019}, and UNIONS \citep{Gwyn2025,Smith2023}. Looking ahead, the Rubin Observatory’s Legacy Survey of Space and Time (LSST) should, in principle, be able to detect every MW satellite in the Southern Hemisphere \citep{Tsiane2025}.

\begin{figure}
\centering
\includegraphics[width=\linewidth]{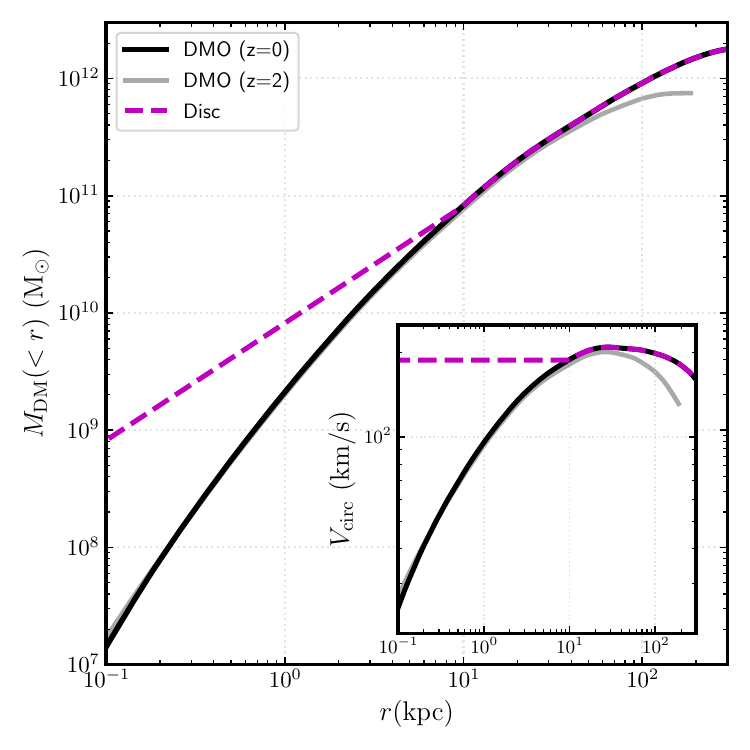}
\caption{
Enclosed mass profile and circular velocity of the  Aq-A-L1 dark matter halo, for the DMO and Disc potential models.  Results at $z=0$ are shown in black and at $z=2$ in grey. 
  }
 \label{fig:host}
\end{figure}

Given these optimistic observational prospects, it is timely to provide robust $\Lambda$CDM predictions for the \textit{complete} satellite population of the MW. Progress in addressing this question has come from both zoom-in hydrodynamical simulations of MW-mass halos \citep[e.g.][]{Wetzel2016,Sawala2016,Garrison-Kimmel2019,Grand2021,Applebaum2021,Font2021,Engler2021}
 as well as semianalytical galaxy formation models \citep[e.g.][]{Benson2002,Bullock2000,Font2011}. 
 These studies have shown that, once key physical processes such as reionization and stellar feedback are included, the bright end of the MW satellite luminosity function can be readily reproduced.

 However, current hydrodynamical simulations are unable to reach the resolution required to follow the formation of the faintest galaxies, precluding direct simulation studies of the ultra-faint regime. Semianalytical models grafted onto high-resolution cosmological dark matter halo simulations seem at present the only practical way forward to make robust predictions that can be contrasted with observations \citep[see e.g.][]{Newton2018,Manwadkar2022,Ahvazi2025}.

A further challenge arises because finite numerical resolution limits the accuracy with which cosmological simulations can follow subhalos. Some of these systems are subject to enormous tidal forces and large tidal mass losses. This leads to the disruption of many subhalos, especially those which enter the halo of the Milky Way quite early, or are on orbits with small pericentric radii  \citep[see, e.g.,][]{vanDenBosch2018}.

 It is now clear that such disruption is artificial,  because in $\Lambda$CDM the cuspy central density profile of subhalos should prevent their total tidal disruption \citep[e.g.][]{Penarrubia2010,Errani2020}. The effect is not subtle. \citet{Santos-Santos2025} estimate that even in the case of the Aq-A-L1 halo \citep[the highest resolution cosmological N-body simulation of a MW-sized $\Lambda$CDM halo ever performed, with particle mass $1.7\times 10^3$ M$_\odot$;][]{Springel2008} about {\it half} of all subhalos deemed capable of hosting a luminous satellite do not survive as self-bound systems until the present time. It is therefore imperative to correct for this artificial disruption in order to provide robust predictions, especially in the ultra-faint regime.

 In a previous paper \citep[][hereafter \citetalias{Santos-Santos2025}]{Santos-Santos2025},  we reported on a first attempt to identify and track all luminous subhalos in the Aq-A-L1 run. Using this model, we were able to predict the total number of satellites that were {\it ever} accreted into the halo of a galaxy like the Milky Way, as well as its dependence on the redshift of reionization. As mentioned above, that work showed that about half of all luminous satellites in Aq-A-L1 are ``orphans''  at present (i.e., their subhalos have not survived to the present as self-bound entities), which emphasizes the critical role that tidal stripping plays in shaping the satellite population.

 The intent of the present study is to extend  \citetalias{Santos-Santos2025}'s work to include a detailed treatment of tidal stripping, based on ``tidal tracks'' derived from idealized simulations designed specifically to circumvent the problems of limited numerical resolution \citep{Errani2021,Errani2022}.  The model tracks all surviving and orphan luminous subhalos and implements analytic prescriptions to estimate  dark matter and stellar mass loss due to tides. Our model improves upon recent semianalytic studies of the ultra-faint population in MW-mass halos which omit tidal stripping effects \citep[see][]{Manwadkar2022,Ahvazi2025}, and therefore cannot predict accurately the present day orbits and structure of satellite galaxies nor track the evolution of their luminosities and sizes.

Using  our model, we are able to present predictions for the abundance, luminosities, sizes, radial distribution, orbital parameters, densities, and velocity dispersions of the \textit{complete} MW satellite population at $z=0$.
In particular, we quantify the number of ultra-faint MW satellites that remain ``missing''  from the current inventory and place constraints on their structure, orbits, and radial distribution.

Our study also considers the effect of a central baryonic disc with properties similar to that of the MW.  Previous work has suggested that such a disc could have a strong effect, enhancing the overall importance of tidal stripping on the satellite population \citep[e.g.][]{Kelley2019,Richings2020}. We note, however, that those conclusions were based on studies solely of the ``surviving'' subhalo population in N-body simulations, and may thus change once the full population of satellites, including orphans, is considered.

Our predictions are particularly timely given the recent discovery of dozens of extremely compact ultra-faint systems (UFCSs) whose ambiguous properties  blur the line between dwarf galaxies and star (globular) clusters \citep[e.g.][]{Smith2024,Cerny2026}. We examine here two scenarios, which explore whether UFCSs are either the heavily stripped remnants of once more luminous dwarfs, or  ``microgalaxies'' which were born compact and have survived relatively unscathed until today due to the high density of their dark matter halo cusp  \citep{Errani2024b}.

The paper is organized as follows. Section~\ref{sec:methods} describes the cosmological N-body simulation used in this work, Aq-A-L1,  and  the GALFORM galaxy formation model. In particular, Sections~\ref{sec:TT} and \ref{sec:StelStrip} present the tidal stripping model and the results of their application on the Aq-A-L1 satellite system. Sec.~\ref{SecResults} discusses the impact of stripping on the distribution of satellite luminosities and radial distances in Aq-A-L1 (Sec.~\ref{sec:AqResults}). The remaining subsections present our predictions for  MW satellites at $z=0$, including the luminosity function,  sizes,  Galactocentric distances, central densities, and velocity dispersions. Finally, Sec.~\ref{sec:conclu} summarizes our findings and conclusions.

\section{Methods}\label{sec:methods}

\subsection{The Aquarius-A dark matter halo}\label{sec:AqA}

We use the Aquarius-A (Aq-A) dark-matter-only (DMO) simulation of a MW-mass halo from the Aquarius project \citep{Springel2008}. The Aq-A halo was run with the GADGET-3 code assuming a flat $\Lambda$CDM cosmology with WMAP-5 parameters
 \citep[$\Omega_{\rm m}=0.25$; $\Omega_{\Lambda} = 0.75$;
$\Omega_{\rm bar} = 0.045$; $H_0 = 100\, h$ km s$^{-1}$ Mpc$^{-1}$;
$\sigma_8 = 0.73$; $h = 0.73$;][]{Komatsu2009}
at 5 different levels of resolution.  In this work we focus on the highest-resolution level, L1, with adopts a particle mass  of $m_p=1.7\times10^{3}$ M$_\odot$ and Plummer-equivalent gravitational softening length of $\epsilon=20.5$ pc.

These characteristics make Aq-A-L1 the highest resolution simulation of a MW-mass halo available, with $\sim1.5$ billion particles within its virial radius \citep[see also][]{Stadel2009}.
This particle mass ensures that all halos in the simulation capable of forming stars according to our ``critical mass'' assumption (see next subsection) are well resolved.

Halos and bound substructures or ``subhalos'' were identified following the friends-of-friends \citep[FOF,][]{Davis1985} and SUBFIND \citep{Springel2001} algorithms, requiring a minimum of 20 particles to identify a subhalo at any given time. The high resolution of Aq-A-L1 enables the identification of up to 4 levels of embedded substructure, as expected from the nature of hierarchical structure formation in $\Lambda$CDM.

 The Aq-A-L1 host halo has, at $z=0$, a virial\footnote{Virial masses are measured within a sphere with mean density 200 times the critical value for closure. Virial quantities are identified with a “200” subscript.} mass and radius of  $M_{200} =1.8\times 10^{12}$ M$_\odot$ and $r_{200}=245.7$ kpc, respectively, and a concentration of $c_{\rm NFW}=16.1$. Numerical convergence of the host density profile and subhalo mass function across resolution levels has been demonstrated in \citet{Springel2008,Navarro2010}.  Aq-A shows steady growth in virial mass without major late-time mergers. Although the virial mass of the halo changes, the inner mass profile is established early and changes very little thereafter.

We show the circular velocity ($V_{\rm circ}$) and enclosed mass profiles at $z=0$ and $z=2$ in Fig.~\ref{fig:host}.  These are computed using all particles assigned to the host halo by SUBFIND. We shall consider below the DMO mass profile (as read directly from the simulation at $z=0$) and a ``Disc'' case, where, in order to simulate the deepening of the potential well expected from the assembly of the baryonic component of the galaxy, we assume that the circular velocity is constant inside a radius, $r_0$, chosen to represent the galaxy radius.  We adopt $r_0=10$ kpc, a value roughly consistent with the region of the Milky Way where the circular velocity profile is observed to be reasonably flat \citep{Lian2024}.

We stress that this "Disc" model is a spherically-symmetric modification of the host potential intended to approximate the deepening of the potential well due to the presence of a central baryonic component, which we believe is the feature that most strongly affects satellite tidal evolution.  It does not capture the flattening of the potential, inclination-dependent disc shocking, secular disc growth, or presence of a bar/bulge component.

\subsection{The GALFORM semianalytic model}\label{sec:galform}
Dark matter halos and subhalos are populated with luminous  galaxies according to the Durham  semianalytic galaxy formation model GALFORM \citep{White1991,Kauffmann1993,Cole1994,Cole2000,Lacey2016}.  GALFORM implements detailed prescriptions for all physical processes relevant to galaxy formation and evolution: shock-heating and radiative gas cooling in halos, star formation in galaxy discs and in starbursts, galactic black hole growth, feedback from stellar evolution, supernovae and active galactic nuclei, chemical enrichment of stars and gas, reionization and the photoionization of the intergalactic medium, among others.

The free parameters of the model have been tuned primarily to reproduce the observed $b_J$ and $K$-band galaxy luminosity functions in the local Universe, as well as the local HI mass function. The model successfully reproduces many other observables to which it was not calibrated, such as the UV-luminosity function of high-redshift JWST galaxies \citep{Cowley2018,Lu2025} and the overall properties of the MW satellite population \citep{Benson2002,Bose2018}. For details on the calibration  we refer the reader to the references above.

\begin{figure}
\centering
\includegraphics[width=\linewidth]{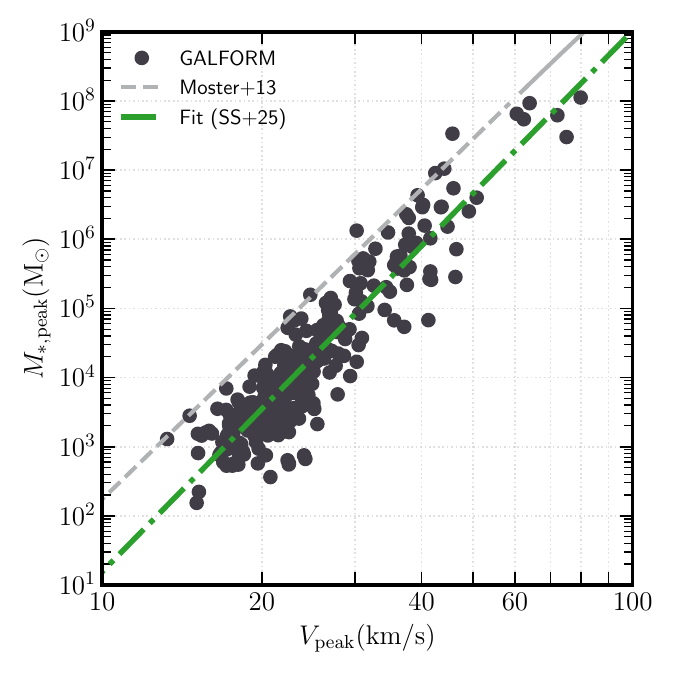}
\caption{   $M_{*,\rm peak}$-$V_{\rm peak}$  relation predicted by GALFORM for Aq-A-L1 satellites at infall time. A grey dashed line shows an extrapolation of the \citet{Moster2013} abundance-matching model.  A green line indicates a power-law fit to the data (see \citetalias{Santos-Santos2025}).}
 \label{fig:Mstvpk}
\end{figure}

\begin{figure*}
\centering
\includegraphics[width=0.47\linewidth]{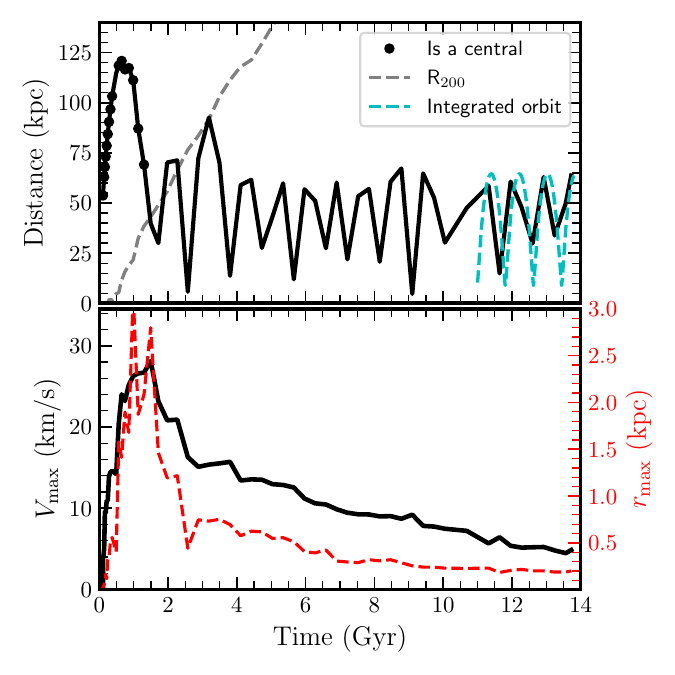}
\includegraphics[width=0.52\linewidth]{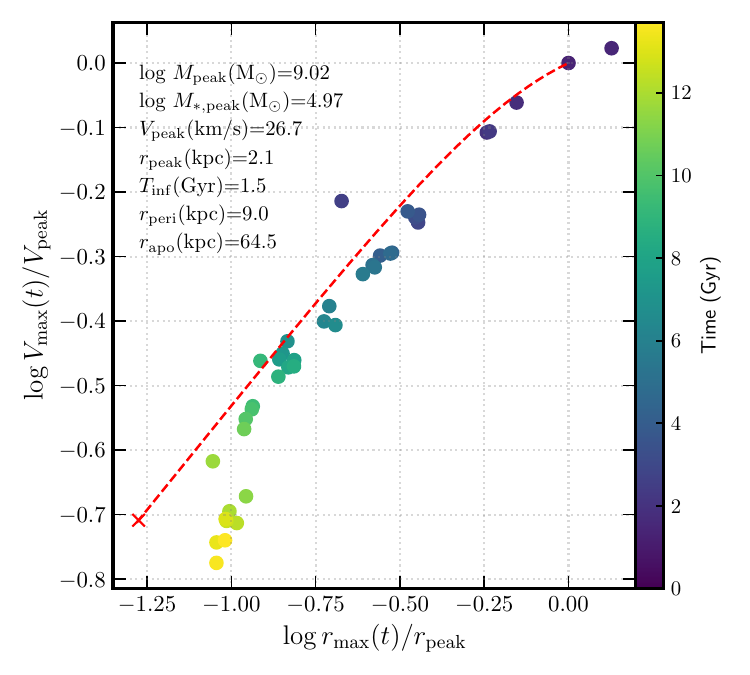}
\caption{ 
Evolution of an example type-1 (``surviving'') Aq-A-L1 satellite.  \textit{Left:} Radial distance from the host halo (top panel), and $V_{\rm max}$, and $r_{\rm max}$ as a function of time (bottom panel).  The grey dashed line in the top panel shows the evolution of the host's $r_{\rm 200}$. The period of time before infall, when the galaxy is considered a central,  is marked with circles. The dashed cyan line shows the orbit integrated backwards in the gravitational potential of Aq-A-L1 from the $z=0$ satellite position and velocity. \textit{Right:} Evolution of $V_{\rm max}$ versus $r_{\rm max}$ for the same satellite,  normalized to their corresponding values at ``peak time'' (i.e., just before infall).   Circles  correspond to values as measured directly by SUBFIND, colored by the age of the Universe at each given timestep.  The darkest point at $(0,0)$ corresponds to the initial ``peak'' (infall) time.   The red dashed line shows the prediction from the tidal track.  The final $z=0$ predicted value is marked with a cross.  Quantities of interest for this example satellite are listed in the legend.
  }
 \label{fig:ttex}
\end{figure*}

\subsubsection{The ``critical'' halo mass for galaxy formation}
\label{SecCritMass}

In the version of GALFORM used here (as well as in \citetalias{Santos-Santos2025}),  gas cooling and subsequent star formation proceed only in halos that exceed a critical mass, $M_{\rm crit}(z)$, computed following the model of \citet{Benitez-Llambay2020}. Before reionization,   $M_{\rm crit}(z)$ is set by the virial halo temperature at which atomic hydrogen can cool (roughly $T_{200}\sim10^4$ K), which corresponds roughly to a peak maximum halo circular velocity of $V_{\rm peak}\sim15$ km/s \citep{Okamoto2009}. $M_{\rm crit}(z)$ decreases with increasing redshift, reaching, for example,  $M_{\rm crit}\sim 10^{7.5}\, M_\odot$, at $z=20$.

After reionization, the critical mass is defined as the halo mass above which photoheated gas cannot remain in hydrostatic equilibrium and thus collapses to the centre of the halo. This  yields a present-day value of  $M_{\rm crit}(z=0)\sim 5\times10^{9}$ M$_\odot$.  (See Fig.~ 2 in \citetalias{Santos-Santos2025} for details on the redshift dependence of $M_{\rm crit}$.) We assume hereafter a redshift of reionization\footnote{For details on the impact of changing this parameter see  the Appendix in \citetalias{Santos-Santos2025}} of $z_{\rm rei}=6$.

GALFORM assumes that the bulk of star formation takes place only  in ``central'' subhalos (i.e., the main subhalo of each FoF grouping). Once a system ``infalls'' and becomes a satellite of another halo, the version of GALFORM used here assumes instantaneous ram pressure stripping of the diffuse gas. From that point on,  no new gas is allowed to cool or accrete onto that subhalo, and only pre-cooled gas can fuel subsequent star formation.  The model does not include tidal stripping (of either dark matter or stellar mass), so additional prescriptions are necessary to estimate realistic stellar masses and sizes of satellites that can be compared to observational data (see Sec.~\ref{sec:TT} below).

These prescriptions lead to a sharp transition between halos that host luminous galaxies and those that remain dark. The transition depends mainly on the value of $V_{\rm peak}$, the highest circular velocity achieved by a halo in its past history. About half of all subhalos with $V_{\rm peak}\sim 20$ km/s harbour luminous galaxies, a fraction that increases to $100\%$ at $V_{\rm peak}\gtrsim 28$ km/s, and which drops to zero at $V_{\rm peak}<13$ km/s (see Fig.~3 in \citetalias{Santos-Santos2025}).

\subsubsection{Surviving and orphan satellites}

GALFORM tracks all luminous subhalos, including those that are artificially disrupted by tidal forces. These orphan satellites are referred to as ``type-2s'' in GALFORM nomenclature,  while ``type-1s''  correspond to satellite galaxies that survive all the way to $z=0$. When a satellite artificially disrupts and disappears from the SUBFIND catalogues (when it drops below 20 particles), GALFORM continues to track it by following the position and velocity of its most-bound particle.  Although approximating the disrupted satellite as a point mass is a crude simplification, we have checked that it provides an adequate description of the orbital evolution, as most orphan systems are very low mass and therefore experience negligible dynamical friction. 

Galaxy mergers between subhalos and the central subhalo are included in GALFORM by computing merging timescales between subhalos and the host halo, following a modification of the analytical Chandrasekhar dynamical friction formula \citep[Eq.~6 in][]{Simha2017}. This affects only the most massive subhalos, and has little effect on the population of low-mass subhalos that host the majority of satellites.

\begin{figure*}
\centering
\includegraphics[width=\linewidth]{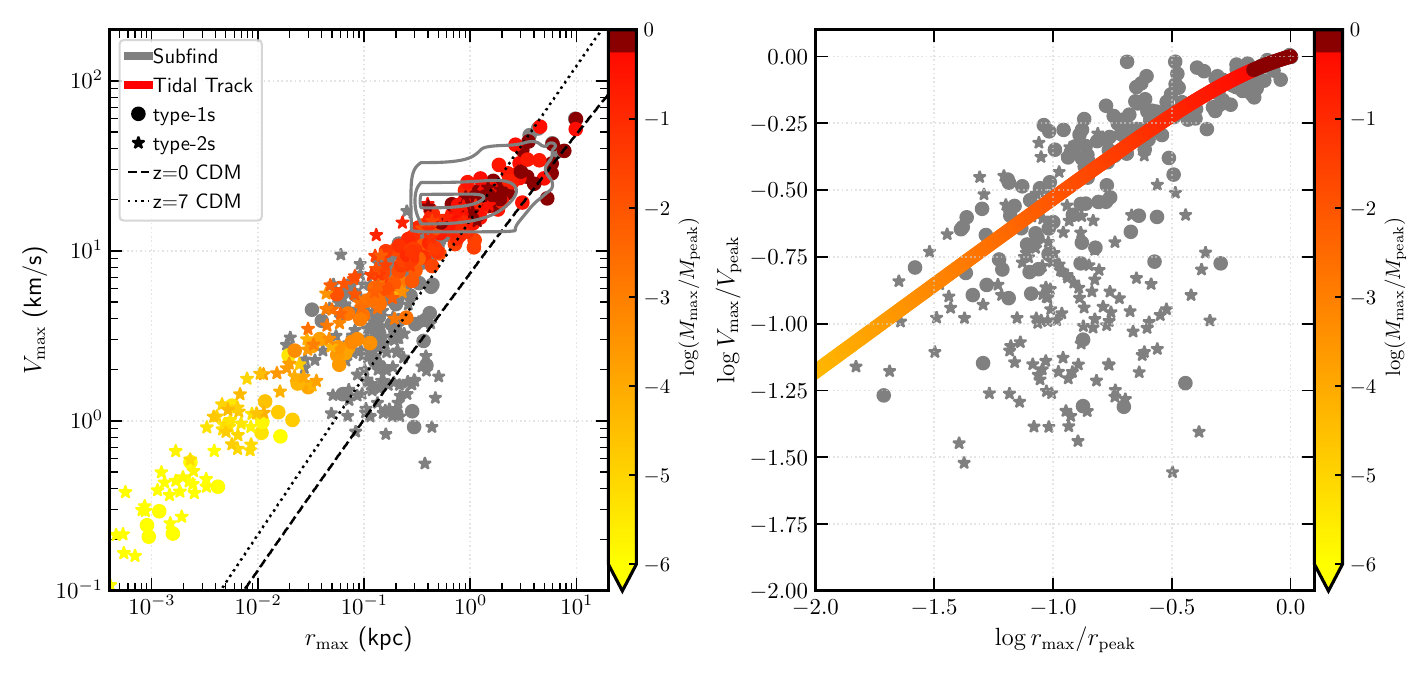}
\caption{Subhalo characteristic radii and velocities ($r_{\rm max}$ and $V_{\rm max}$) of Aq-A-L1 satellites at $z=0$, according to the SUBFIND halo finder catalogue, versus their values as predicted by the tidal tracks of \citet{Errani2021}. \textit{Left:} $V_{\rm max}$ vs $r_{\rm max}$ relation in physical units. Grey symbols (circles for type-1s and stars for type-2s) show the $z=0$ values as measured by  SUBFIND.  In the case of type-2s, values correspond to the last timestep when the object was resolved  by the halo finder before artificial disruption.  
Colored symbols show the corresponding $z=0$ predictions for those same Aq-A-L1 satellites following the tidal track framework.  Symbols are colored according to the
 fraction of self-bound dark matter remaining in the subhalo by $z=0$,  $\rm log(M_{\rm max}/M_{\rm peak})$. The darkest shade of red corresponds to a mass loss of less than $50\%$. The grey contours indicate the location of the peak values  ($V_{\rm peak}$ and $r_{\rm peak}$)  at infall. For reference, a dashed (dotted) black line shows the $z=0 (7)$ $V_{\rm max}$-$r_{\rm max}$ relation expected in $\Lambda$CDM, assuming the average mass-concentration at each redshift  \citep{Ludlow2016}.
\textit{Right:} Same as the left panel but normalized to the peak values (at infall).  Datapoints indicate the change in $V_{\rm max}$ and $r_{\rm max}$ from their peak time values to $z=0$ (or, for type-2s, to the last timestep before  the object was disrupted). The colored  line shows a tidal track, which by definition starts at $(0,0)$ in this plane.  Colors follow the same colorbar as in the left panel.  
}
 \label{fig:ttphys}
\end{figure*}

\subsection{Aq-A-L1's unabridged satellite population}

We hereafter define as satellites all luminous galaxies (including orphans) found within $300$ kpc of the Aq-A-L1 host halo at $z=0$. This results in $320$ satellites when assuming $z_{\rm rei}=6$.  For results for other reionization redshifts, see appendix A in \citetalias{Santos-Santos2025}; for example, satellite number is reduced to $240$ for $z_{\rm rei}=10$. 

We showed in \citetalias{Santos-Santos2025} that orphan satellite fractions increase at lower simulation resolution, and that orphan satellites inhabit subhalos with systematically earlier infall times as well as smaller pericentric and apocentric distances than surviving satellites. As stated above, orphans contribute at least half of the total number of luminous satellites accreted onto the Aq-A halo, even at the extremely high resolution of Aq-A-L1. 

GALFORM tracks the time of infall of each satellite, $T_{\rm inf}$, defined as the age of the Universe when a satellite first crosses the evolving virial radius of the host. For a satellite, the time when it reaches its ``peak''  circular velocity roughly coincides with the time at which its bound mass also peaks, and usually occurs just before infall. GALFORM also records the peak stellar mass of a satellite,  $M_{*,\rm peak}$, its stellar half-light radius, $r_{\rm half, peak}$,  as well as the radius where the peak circular velocity is reached, $r_{\rm peak}$.

\subsubsection{Satellite stellar masses}

Applied to Aq-A-L1, GALFORM  yields a power-law-like relation linking  $M_{*,\rm peak}$ and $V_{\rm peak}$,  as shown in Fig.~\ref{fig:Mstvpk}.
The green line marks the best-fitting power-law fit , which may be expressed as 
$M_{*,\rm peak} = 9 \times 10^8 (V_{\rm peak}/ 100$ km s$^{-1}) ^{7.8}$ M$_\odot$ over the range $15<V_{\rm peak}/$km/s $<80$
 (\citetalias{Santos-Santos2025}).  Compared with an extrapolation of the  abundance-matching model of \citet{Moster2013} GALFORM galaxies are systematically less luminous, at fixed $V_{\rm peak}$. One feature of the GALFORM results is that  the satellite stellar mass function at infall peaks at $10^{3.5}$ M$_\odot$, with essentially no satellites predicted to have $M_{*,\rm peak}\lesssim 500\, M_\odot$  (see Fig.~4 in \citetalias{Santos-Santos2025}).

\begin{figure*}
\centering
\includegraphics[width=0.65\linewidth]{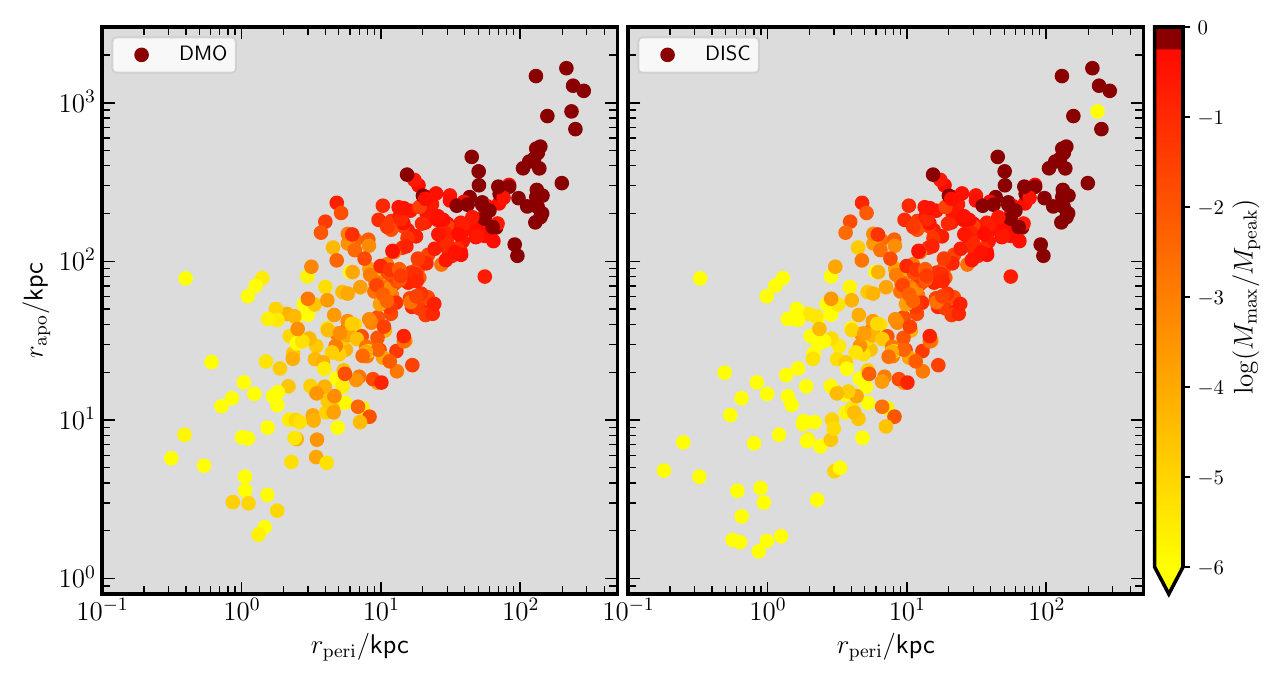}
\includegraphics[width=0.345\linewidth]{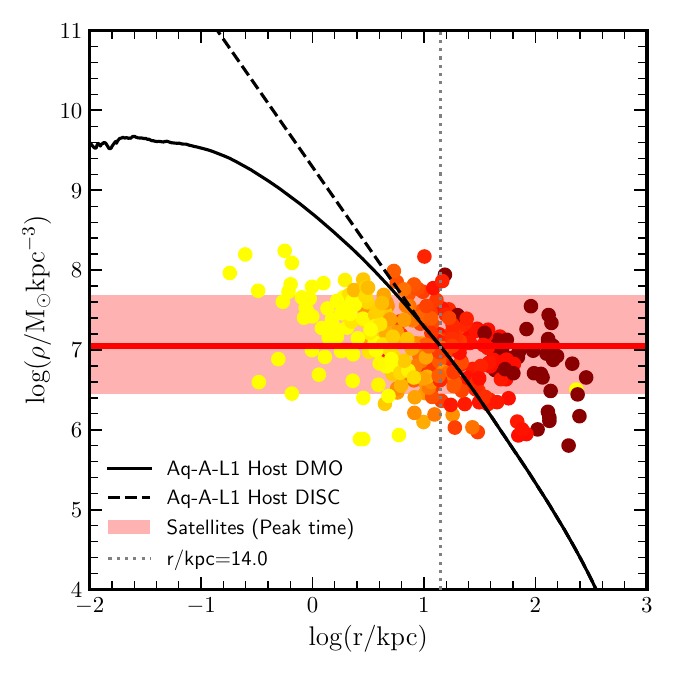}
\caption{ \textit{Left and Middle panels:} Pericentric vs apocentric distances computed for Aq-A-L1 satellites using  the DMO (left) or Disc (middle) potentials.  Points are colored by the fraction of dark matter mass bound to the subhalo by $z=0$, after stripping, computed using the tidal track model described in Sec.\ref{sec:TT}.
\textit{Right:} Points show the mean dark matter density within $r_{\rm peak}$ of Aq-A-L1 satellites at infall time, versus their pericentric distance $r_{\rm peri}$.   Points are colored by mass loss, as in the other panels.  The median of these values is shown as a  horizontal red  line  and the shaded region represents the 25-75th percentiles. This is compared to the mean enclosed dark matter density profile of the host halo in the  DMO (solid) and Disc (dashed) potential cases.  A vertical line marks the radius at which the host density profile and  the  median subhalo density intersect.  Significant mass loss is expected  for subhalos with pericentres for which $\bar \rho_{\rm sat}(<r_{\rm peak})<\bar \rho_{\rm host}(r=r_{\rm peri})$, i.e., those that reach radial distances below $\lesssim15$ kpc.
  }
 \label{fig:massloss}
\end{figure*}

\subsubsection{Satellite orbits}

Since the snapshot spacing of the simulation is not frequent enough to resolve satellite orbits accurately, we estimate pericentric and apocentric distances by integrating their orbits back in time using the position and velocity of each satellite at $z=0$ (as measured in the DMO simulation)\footnote{For orphans (type-2 satellite galaxies), this phase-space information
is given by the most-bound particle.}. The host potential is assumed to remain static during the integration.  Specifically, we integrate $2$ Gyr back in time, which is sufficient to track the orbits of most satellites over several pericentric passages.  Orbital pericentric, $r_{\rm peri}$, and apocentric, $r_{\rm apo}$, distances are computed for both the DMO and Disc host potentials.

For simplicity, we assume that the addition of the disc does not alter the present-day position and velocities\footnote{We recognize that rerunning the simulation with the Disc potential could enhance the tidal field and tidal stripping, leading to the artificial disruption of some systems by $z=0$ that otherwise survive in the DMO case. This difference, however, is  implicitly corrected for in our model,  which assumes that any disappearance is artificial and tracks the evolution of all orphan galaxies  (see \citetalias{Santos-Santos2025} and Sec.\ref{sec:galform}). } nor the subhalo merger history, which only affects the most massive subhalos. Thus, before stripping,  the initial  $z=0$ satellite population is assumed to be the same in both potentials, although they will be subject to different tidal stripping corrections, which we describe next.
 
\subsection{Tidal stripping model: dark matter} \label{sec:TT}

The mass loss and structural evolution of dark matter subhalos are modelled using the tidal track framework of \citet{Errani2021}.  Using high-resolution $N$-body experiments, these authors showed that tides acting on a $\Lambda$CDM  subhalo (modeled as an NFW density profile) lead to a bound remnant with asymptotic properties set solely by the subhalo's initial structure and the density of the host halo at pericentre.

The model empirically describes the evolution of a subhalo's structure,  revealing a well-defined trend between the mass loss due to tides (measured via the evolving maximum circular velocity $V_{\rm max}$),  and the change in characteristic size $r_{\rm max}$. (Tidal effects can also be measured also by using as a parameter the total mass inside $r_{\rm max}$, $M_{\rm max}=V_{\rm max}^2 \, r_{\rm max}/G$.) 
When normalized to their initial values,  the tidal evolution of every subhalo in CDM follows roughly the same single line (or tidal track)  in the $V_{\rm max}(t)/V_{\rm peak}$ vs $r_{\rm max}(t)/r_{\rm peak}$ plane.  At fixed $r_{\rm peri}$, the orbital eccentricity affects only the rate at which a subhalo moves along the track; the higher the eccentricity the slower tidal stripping proceeds at fixed pericentric distance. The universality of subhalo tidal evolution is a consequence of the universal density profile of CDM subhalos and of its associated binding-energy distribution. 

For each Aq-A-L1 satellite, we numerically compute its tidal evolution using the public code provided by \citet{Errani2021}\footnote{\url{https://github.com/rerrani/tipy}}. This requires as input the satellite’s orbital parameters: $r_{\rm peri}$,  $r_{\rm apo}$; its peak (infall) quantities,  $V_{\rm peak}$,  $r_{\rm peak}$, as well as the 
circular velocity profile of the host at pericentre, $V_{\rm circ}(r=r_{\rm peri})$. The code returns the bound mass,  $M_{\rm max}$, characteristic radius, $r_{\rm max}$, and maximum circular velocity, $V_{\rm max}$ as a function of time. With these quantities we then compute, as a function of time, the density profile of the bound remnant, which is well approximated by a truncated NFW profile \citep[see eqs.7-9 in][]{Errani2021}.

Fig.~\ref{fig:ttex} illustrates the tidal evolution of an example Aq-A-L1 type-1 satellite. The left-hand panel shows the evolution of the satellite's radial distance with respect to the host centre (upper panel),  as well as the evolution of $V_{\rm max}$ and $r_{\rm max}$ (lower panel). The grey dashed line in the upper panel shows the evolution of the host halo's virial radius.  Black circles  identify the times when the satellite was considered a central  (i.e., not yet a satellite of the Aq-A-L1 host halo).

This satellite is accreted into the host halo at early times, $T\sim1.5$ Gyr, and settles into a relatively short-period orbit with small pericentric and apocentric distances. The satellite is able to complete more than $17$ orbits by $z=0$, as inferred from our orbit integration scheme (see dashed cyan line).  After infall, both $V_{\rm max}$ and $r_{\rm max}$ decline steeply after each pericentric passage. From an initial $V_{\rm peak}=26.7$ km/s, the satellite's $V_{\rm max}$ is reduced to roughly $4.9$ km/s at $z=0$.

The right-hand panel shows the subhalo's evolution in the $\log V_{\rm max}/V_{\rm peak}$–$\log r_{\rm max}/r_{\rm peak}$ plane, compared with the tidal track (red line).
The circles indicate measurements taken from the SUBFIND catalogue at different times. Note that the subhalo follows the tidal track rather accurately at the beginning, but eventually deviates once the number of bound particles becomes insufficient to adequately resolve the structure.  This occurs at $T\approx 10$ Gyr, when $V_{\rm max}=7.5$ km/s and the bound particle number drops to $3150$.  This is in good agreement with \citet{Errani2021},  who argued that at least $\sim3000$ particles within $r_{\rm max}$ are required to reliably follow CDM subhalo tidal evolution.  

This example demonstrates how even the highest resolution MW halo simulations currently available still suffer from numerical resolution limitations,  so that tracking subhalo structural properties can become unreliable even well before subhalos are disrupted. Our analysis will therefore use tidal tracks to estimate the evolution of the structural parameters of subhalos after infall rather  than the results from the N-body simulation. For the satellite shown in Fig.~\ref{fig:ttex}, the $z=0$ structure of the subhalo indicated by the tidal track is shown by the red cross in the right-hand panel.

 \subsubsection{Aq-A-L1 subhalos after tidal stripping}\label{subsubsec:massloss}

 We apply the tidal track model to Aq-A-L1 and track the evolution of the subhalo structural parameters $r_{\rm max}$ and $V_{\rm max}$   of all luminous subhalos in the DMO and Disc host potentials.  The results are shown in the left panel of  Fig.~\ref{fig:ttphys}, where each satellite is colored by the remaining self-bound mass fraction (circles for type-1s and stars for type-2/orphans). We show results here for the Disc potential, although  results are quite similar for the DMO case (see Fig.~\ref{fig:massloss} below).

The grey symbols in Fig.~\ref{fig:ttphys}  indicate the results from SUBFIND at $z=0$. For type-2s,we use the values corresponding to the last time when the satellite was identified in the simulation before artificial disruption. We can   see that many of them have fallen below the tidal track traced by the colored symbols, a clear indication of the effects of limited numerical resolution \citep[see the discussion in the Appendix of][]{Errani2021}.

Grey contours indicate the loci in this plane for satellites at infall. As satellites tidally evolve, their positions move away from the initial contour region, roughly along diagonal lines which, individually, trace the tidal track. Aq-A-L1 satellites suffer a wide range of tidal mass loss; some systems are predicted to retain less than $10^{-6}$ of their infall mass ($M_{\rm peak}$, see colorbar), whereas others survive more or less unscathed to $z=0$. 

The right-hand panel of Fig. ~\ref{fig:ttphys} is analogous to the left-hand panel, but the values of $V_{\rm max}$ and $r_{\rm max}$  have now been  normalized to their values at infall.  The coloured line shows the universal tidal track expectation, with colors again varying according to the remaining self-bound mass fraction. While some satellites lie close to the track, many fall below it, deviating from the predicted track.  As expected, the largest deviations are found amongst type-2 systems.

\begin{figure}
\centering
\includegraphics[width=\linewidth]{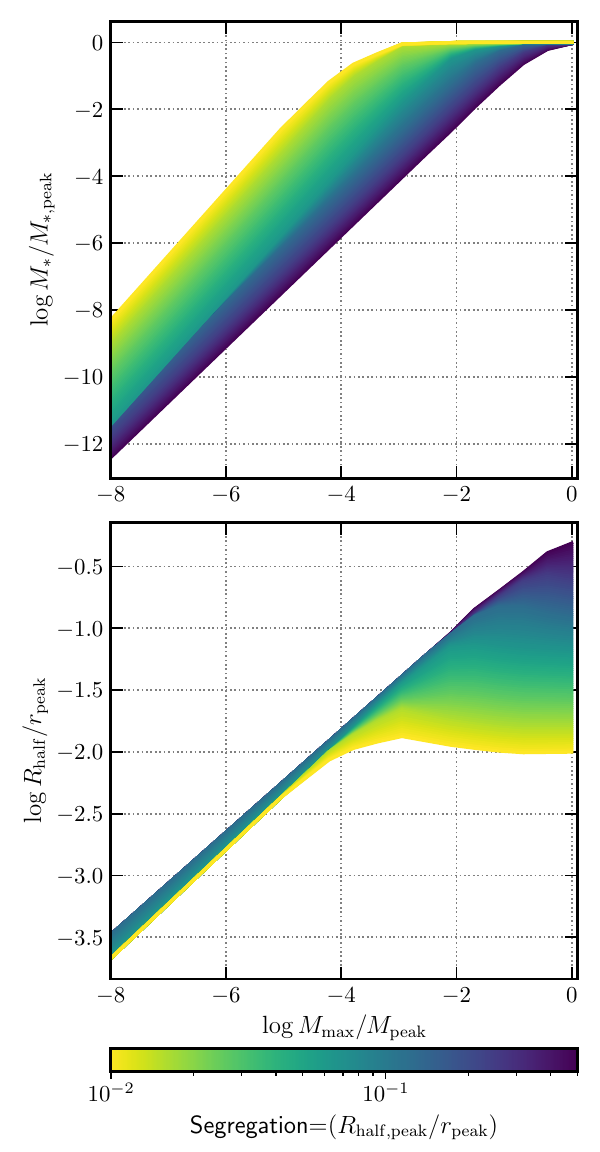}
\caption{Stellar tidal tracks from \citet{Errani2022} used in this work. Lines of constant color indicate the track of a model with fixed initial radial segregation values, where  segregation is defined as $S= R_{\rm half}/r_{\rm peak}$.  The plot shows the change in stellar mass (\textit{top}) and change in size (\textit{bottom}),  relative to their peak values at infall time, as a function of remnant subhalo mass fraction, $\log M_{\rm max}/M_{\rm peak}$ (see  Sec.~\ref{sec:TT}).
  }
 \label{fig:model}
\end{figure}

\begin{figure*}
\centering
\includegraphics[width=1\linewidth]{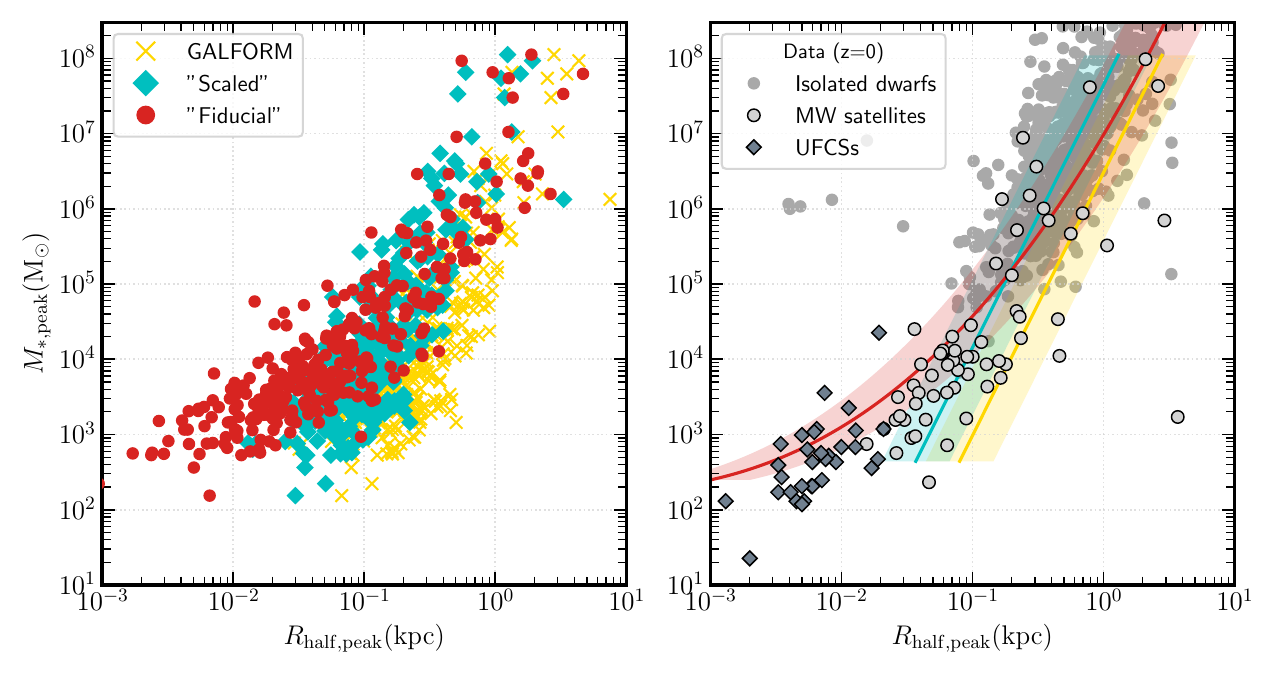}
\caption{
$M_{*,\rm peak}$-$R_{\rm half,peak}$ relations assumed in this work for Aq-A-L1 satellites.  $M_{*,\rm peak}$ values are  predicted by GALFORM  (Fig.~\ref{fig:Mstvpk}), but stellar half-light radii are assigned according to different models.  In the left panel, each color corresponds to a different model. Yellow crosses show $R_{\rm half,peak}$ as predicted by GALFORM.  Cyan diamonds show the "Scaled"  model,  where $R_{\rm half,peak}$
 is a result of scaling down GALFORM sizes by a factor of $3$, resulting in a distribution of approximately 
 fixed segregation of $S=1/15$. Red circles show the "Fiducial"  model, where satellites as compact as the MW Ultra-Faint Compact Satellites are allowed to form in very faint systems (see text for details). The right panel shows with colored lines the average relation of each model with its scatter, compared to observational $z=0$ data (grey symbols, see Appendix \ref{app:data}). 
  }
 \label{fig:MstRh}
\end{figure*}

\subsubsection{DMO vs Disc host potential}

We now compare the difference in tidal mass loss that results from assuming either the DMO or the Disc host potential in the tidal track model. Fig.~\ref{fig:massloss} shows the pericentric and apocentric distances computed for each satellite, colored by the fraction of bound dark matter mass at $z=0$, as in  Fig.~\ref{fig:ttphys}.
The left panel  of Fig.~\ref{fig:massloss} shows results for the DMO potential, while the middle panel shows results for the Disc potential.  

The results are quite similar for both potentials.
With the inclusion of a disc,  only a very small fraction of objects (those with $r_{\rm peri} \lesssim 10$ kpc) show any change in their computed orbital parameters,  and consequently in the amount of tidal mass lost.

Note the strong dependence of mass loss on pericentric distance.  Satellites with $r_{\rm peri}\lesssim 10$-$20$ kpc experience extreme mass loss, and are able to retain less than $0.0001\%$ of their initial mass. On the other hand, the majority of satellites with $r_{\rm peri}\gtrsim20$ kpc retain more than $10\%$ of their initial $M_{\rm max}$. The inclusion of the $10$ kpc Disc (middle panel) does not alter these results appreciably, mainly because relatively few satellites reach such small distances ($<10$ kpc) from the host halo centre.

The right-hand panel of Fig.~\ref{fig:massloss}  illustrates why tidal mass loss changes rather drastically for pericentres around $r_{\rm peri}\lesssim 15$ kpc. The black curves show the mean enclosed density profile of the host in the DMO (solid) and Disc (dashed) potentials. The  symbols show the  mean characteristic density (i.e., $\bar \rho_{\rm peak}=V_{\rm peak}^2/(4\pi G/3)\, r_{\rm peak}^2$) of Aq-A-L1 satellites at infall time versus their corresponding pericentric distances.  As in the left-hand panel, symbols are colored by mass loss. The horizontal red line marks the median value and the shaded band the 25-75th percentile range.

The right-hand panel of Fig.~\ref{fig:massloss}  shows that the median subhalo density at infall is comparable to the host halo's mean enclosed density at $r\sim 15$ kpc, marked with the vertical dotted line. Satellites with pericentres smaller than this radius hence experience especially severe tidal mass loss, because they orbit regions where the mean density of the host exceeds $\bar \rho_{\rm peak}$.  Since the baryonic disc radius in our model is $10$ kpc (i.e. smaller than $15$ kpc), the same comment applies to both the  DMO and Disc potentials.

These results suggest that the presence of a central massive baryonic disc, of size comparable to that observed in our MW Galaxy,  does not have a critical effect on the tidal mass loss of satellites around MW-mass hosts. This appears to contradict previous studies, which have claimed that the MW's disc exacerbates satellite tidal stripping and disruption \citep[see, e.g.,][]{Kelley2019,Graus2019,Bose2020}.  This apparent contradiction arises because the addition of a disc does actually enhance the {\it artificial} tidal disruption of subhalos with small pericentric distance, which may lead to large changes in the number of {\it surviving} subhalos. However, once artificial disruption is corrected for and orphan satellites are taken into account,  our results indicate that the extra impact of adding a central disk on the tidal evolution of most satellites is relatively minor.

We end by noting that our tidal-stripping model does not distinguish between different physical mechanisms of tidal mass loss. For example, the impulsive "disc shocking" experienced by a satellite crossing a geometrically thin baryonic disc is expected to affect a satellite's internal structure differently from the more gradual, orbit-averaged stripping it would experience on an eccentric orbit within a spherical NFW-like potential. Our spherical "Disc" potential is designed to capture what we consider the dominant effect of a central baryonic component, i.e., the deepening of the potential well, rather than these more localized tidal mechanisms, which may be significant for individual satellites but are not expected to change our conclusions at the population level. Our work represents the first attempt to model tidal stripping accurately and self-consistently across the full cosmological satellite population of a MW-mass halo at high resolution;  a more complex treatment of disc/bulge-specific tidal channels, building on the framework introduced here, is left to future work.

\subsection{Tidal stripping model: stars}\label{sec:StelStrip}

The effects of tidal stripping on the stellar component of satellite galaxies may be estimated using the model described by \citet{Errani2022}. This model assumes that galaxies are dark matter dominated and that their stars follow an exponential stellar density profile\footnote{More explicitly, we adopt the model corresponding to eq.~13 in \citet{Errani2022}, with $\alpha=\beta=3$. See that paper for further details.}  embedded in an NFW halo. These are assumptions particularly well-suited for the evolution of dwarf spheroidal galaxies, which make up the majority of MW satellites.

With these assumptions, the main parameter that controls how stars are stripped is the degree of radial segregation of stars relative to dark matter. We  define the segregation parameter of a satellite galaxy , $S=R_{\rm half}/r_{\rm peak}$, as the ratio between the initial (infall) stellar 2D projected half-light radius, $R_{\rm half}$,  and the characteristic subhalo size $r_{\rm peak}$. Smaller segregation values indicate more deeply-embedded stellar systems.
 
Fig.~\ref{fig:model} shows  our stellar stripping model (adapted from Fig.~F1 from \citealt{Errani2022}), 
showing the expected evolution of stellar mass (upper panel) and half-light radius (lower panel) as a function of the remaining dark matter bound mass fraction, $M_{\rm max}/M_{\rm peak}$. 
Lines of fixed color correspond to the ``tracks''  followed by systems with different initial segregation values, as given by the color bar at the bottom of the figure.

For a given satellite, then, the impact of tides on its stellar content depends mainly on its orbit, which will determine how the dark matter, $M_{\rm max}$, evolves as a function of time, and on $S$, which will then determine how much stellar mass has been lost and how much the satellite has shrunk relative to its original size.  We adopt initial stellar masses as given by GALFORM (Fig.~\ref{fig:Mstvpk}), and discuss  different models for the initial/infall size of satellites next.

 \subsubsection{Satellite size-mass relation}

 The yellow crosses in the left panel of Fig.~\ref{fig:MstRh} show the peak stellar mass vs 2D half-light radii of Aq-A-L1 satellites as predicted by GALFORM. In the right-hand panel, the yellow line shows a power-law fit to these data, compared with observational estimates for dwarf galaxies in the Local Universe (shown in grey; data sources are listed in Appendix~\ref{app:data}). It is clear that GALFORM predicts dwarf galaxies that, at given stellar mass, are systematically too spatially extended compared with observations of isolated dwarfs.

This motivates us to explore alternative prescriptions for assigning initial/infall stellar sizes to our Aq-A-L1 satellites, while maintaining their GALFORM-predicted peak stellar masses. As a first possibility, we consider a model where all GALFORM galaxy sizes are scaled down by a factor of three,  shown with the  cyan diamonds  in the left panel of Fig.~\ref{fig:MstRh}. This results in a narrow distribution of segregation values,  well described by a lognormal with a characteristic value $S=0.07 (\sim 1/15)$ and dispersion of $0.25$ dex; i.e., all satellites have roughly the same initial ratio between effective radius and the characteristic size of its dark matter halo; $S\approx$ constant. 
  The cyan line in the right-hand panel of Fig.~\ref{fig:MstRh} shows a power-law fit to these data, which shows that this ``Scaled'' model is in better agreement with the overall trend seen for isolated dwarfs (the majority of which have $M_*>10^5\,M_\odot$) and for MW satellites. 

As discussed in Sec.~\ref{sec:intro},  deep imaging campaigns (e.g. SDSS, DES, UNIONS)  have recently uncovered a population of UFCSs, or ultra-faint compact MW satellite candidates.  With only a handful of stars within spectroscopic reach in each of these systems, velocity dispersion estimates are highly uncertain, and their dark matter content has not yet been reliably determined. This implies that their classification as either dwarf galaxies or star clusters is unclear. The UFCS population is shown with dark grey diamonds in the right-hand panel of Fig.~\ref{fig:MstRh}.

Motivated by these new discoveries, we consider a second, ``Fiducial'', model, where the initial $M_*$–$R_{\rm half}$ relation deviates from a power-law, bending at low luminosities towards systematically smaller sizes, allowing for the formation of systems initially as compact\footnote{Some semianalytic models suggest that the formation of such systems is indeed possible in LCDM if star formation is driven by molecular H$_2$ cooling \citep{Manwadkar2022,Ahvazi2025}.} as the UFCS population.  The adopted relation, with scatter (a factor of 2 horizontally), is shown with a red line and shaded area in the right-hand panel of Fig.~\ref{fig:MstRh}. Individual values are shown by the red circles in the left-hand panel of the same figure, where we have assigned initial $R_{\rm half, peak}$ values to each satellite's $M_{*,\rm peak}$ by randomly sampling the ``Fiducial'' relation\footnote{The Fiducial relation can be expressed in parametric form, where $R_{\rm half}/{\rm pc} = 10^{4.2\,x}$, $M_V = -13\, (x + 0.25)^{2.5}$,  and $x$ is an auxiliary variable such that $x \in [0,1]$.  This function assumes that all satellites have peak/infall luminosities $M_V \lesssim 0$ (or $M_*\gtrsim 100$ M$_\odot$),  consistent with GALFORM predictions.  For the red line shown in the right-hand panel of Fig.~\ref{fig:MstRh} we have assumed $M_*/L_V=2$.  For each satellite with $M_{*,\rm peak}$ as predicted by GALFORM, we randomly sample an $R_{\rm half}$ value assuming this Fiducial relation follows a lognormal distribution  with a dispersion of $0.3$ dex. The result of one random realization is shown by the red circles in the left-hand panel of Fig.~\ref{fig:MstRh}.}.
 In this case, satellites have a wide range of initial segregation values, with systems becoming more deeply embedded in the cusp of their dark matter halos as their stellar mass decreases.  We shall adopt these two models to set the initial stellar masses and sizes of satellite galaxies assumed in this work prior to undergoing tidal stripping.

 \subsubsection{Tidal evolution of the stellar component of Aq-A satellites}
 
\begin{figure}
\centering
\includegraphics[width=\linewidth]{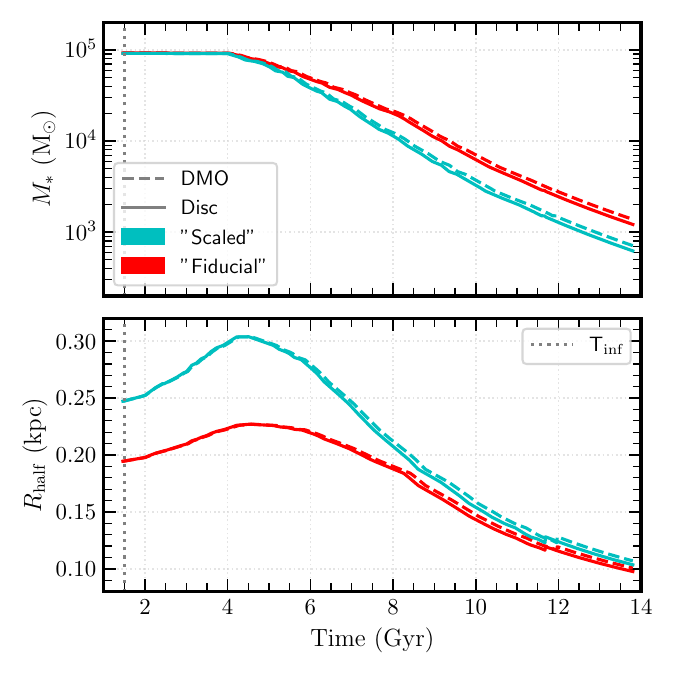}
\caption{ Evolution of the bound stellar mass (top panel) and 2D projected half-light radius (bottom panel) of the example satellite shown in Fig.~\ref{fig:ttex}, according to the stellar stripping model described in Sec.~\ref{sec:StelStrip}. Cyan (red) curves correspond to the "Scaled" ("Fiducial") luminosity-size prescription.  Solid (dashed) lines show results assuming the "DMO" ("Disc") host potentials. The evolution of this example satellite galaxy is shown starting from its infall time, $T_{\rm inf}$, which is marked with a vertical dotted line.
 }
 \label{fig:Lrhalfevo}
\end{figure}

Fig.~\ref{fig:Lrhalfevo} illustrates the tidal evolution of the stellar component of the same satellite chosen to illustrate the dark matter evolution  in Fig.~\ref{fig:ttex}.   Cyan lines in Fig.~\ref{fig:Lrhalfevo} correspond to the Scaled model ($S=0.12$), and red lines to the Fiducial model ($S=0.09$). Solid and dashed curves represent results in the DMO and Disc Aq-A-L1 host potentials, respectively. The vertical dotted line marks the satellite's infall time,  the starting point of the tidal evolution.

The top panel shows the tidal  stellar mass loss. As shown by their segregation values, the Scaled model galaxy is initially slightly larger than the Fiducial one, which implies that the Scaled model loses stars more rapidly. However, by the present day, both models predict that only about $\sim 1\%$ of the original stellar mass remains bound to the satellite.

The bottom panel in Fig.~\ref{fig:Lrhalfevo} shows the evolution of $R_{\rm half}$.  In both models, $R_{\rm half}$ first  increases because the removal of dark matter particles induces stellar expansion as stellar orbits respond to the changing subhalo potential \citep{Penarrubia2008}.
 After some time, once the halo has been stripped down to the edge of the stellar distribution, $R_{\rm half}$ begins to decrease.  The initial expansion phase appears more pronounced in the Scaled model than in the Fiducial one,  while the shrinking phase proceeds similarly in both. By the present time, the satellite has decreased in size by about a factor of two.

The weaker overall  impact of stellar stripping observed in the Fiducial model compared to the Scaled one is a result of their different initial segregation distributions:  in the Fiducial model most satellites have their stars more compactly embedded within the dark matter halo, making their stellar component more resilient to stripping effects. Note as well that tidal losses seem to affect the total stellar mass more than the size of the stellar component. This is a generic property of the \citet{Errani2022} stellar stripping model adopted here.

\begin{figure*}
\centering
\includegraphics[width=0.49\linewidth]{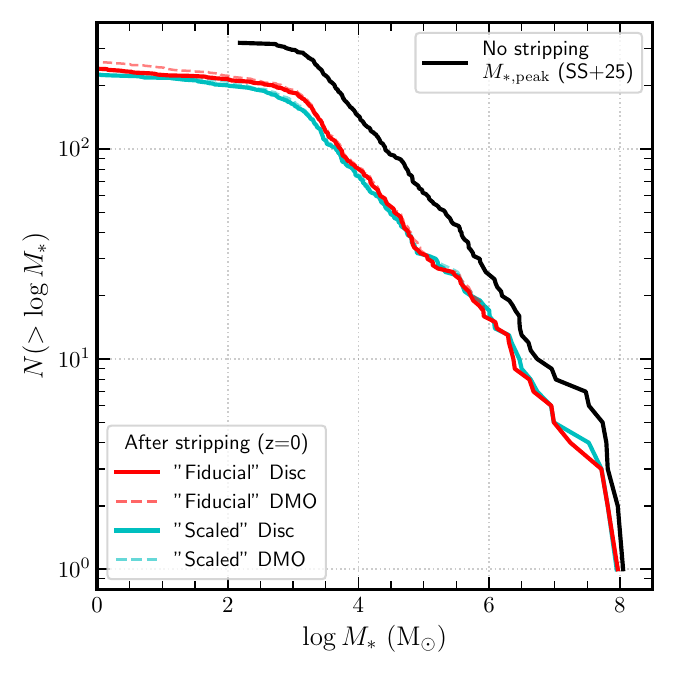}
\includegraphics[width=0.49\linewidth]{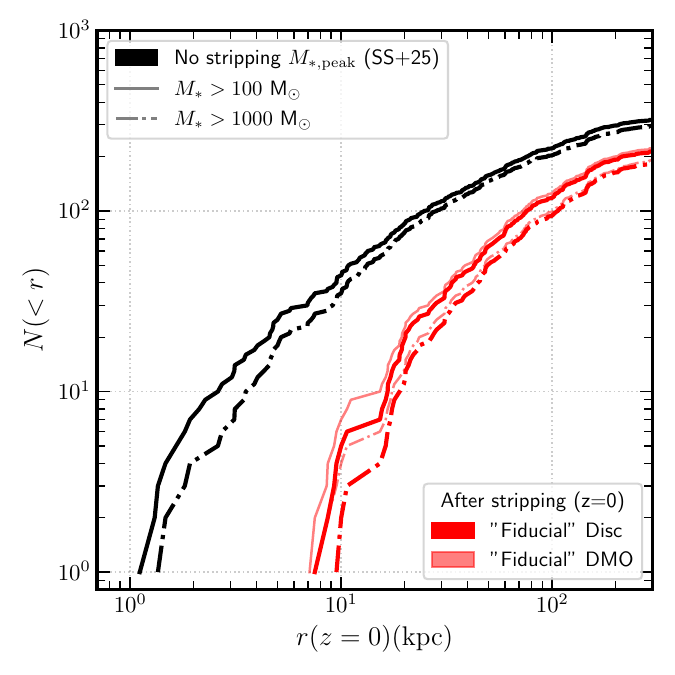}
\caption{ 
\textit{Left:}  Satellite stellar mass function in Aq-A-L1. The black line shows the total (type-1s and type-2s/orphans)   ``unabridged''  stellar mass function of Aq-A-L1 satellites,  at infall time  ($M_{*, \rm peak} $, see fig.4 in \citetalias{Santos-Santos2025}).  
Coloured lines show the predicted $z=0$ functions after accounting for stellar tidal stripping (see details in Secs.~\ref{sec:TT} and \ref{sec:StelStrip}).  
Cyan lines correspond to results assuming the "Scaled" luminosity-size prescription,  while red lines correspond to the "Fiducial" prescription.  Different linestyles illustrate results using the DMO (dashed) or Disc (solid) host potentials.
Note that the coloured $z=0$ functions  include the same total number of satellites as the ``unabridged'' function shown in black,  however some satellites are stripped so severely  that their final stellar masses fall below the plotted range.
\textit{Right:} Cumulative radial distribution of  Aq-A-L1  satellites at $z=0$. Different linestyles show results for satellites with stellar masses above a  certain threshold,  as indicated by the legend.  Black lines show the  radial distribution of all satellites  (type-1s and type-2s),  according to their peak stellar masses (see fig.7 in \citetalias{Santos-Santos2025}).  
Coloured lines show the $z=0$ distributions obtained after applying stellar stripping.
We only show the “Fiducial”  model results,  since radial distances are identical for the “Scaled” case and final stellar masses differ only marginally (see left panel).
 Lighter (darker) red shades correspond to results assuming the DMO (Disc) host potential.
}
 \label{fig:lumfunraddist}
\end{figure*}

\section{Results}\label{SecResults}

\subsection{Tidal effects on the stellar mass function and radial distribution of Aq-A-L1 satellites}\label{sec:AqResults}

The left panel of  Fig.~\ref{fig:lumfunraddist} shows the cumulative stellar mass function of Aq-A-L1 satellites.  The black line shows the predicted function at peak time (i.e., at infall), presented in \citetalias{Santos-Santos2025}.  As shown in Fig.~\ref{fig:Mstvpk}, GALFORM predicts a total of $320$ satellites with a  minimum $M_{*,\rm peak}$ of roughly $10^3\, M_\odot$. This represents the complete, unabridged population of luminous subhalos that ever accreted onto Aq-A-L1 and did not merge with the central galaxy,  including orphan galaxies. Colored lines show the $z=0$ predictions after applying our stellar stripping model assuming either the Scaled (cyan)  or Fiducial (red) mass-radius prescriptions. Solid and dashed coloured lines correspond to results adopting the DMO and Disc host potentials, respectively.

A comparison of the black and coloured lines demonstrates that stellar stripping is an important ingredient of the luminous satellite population of Aq-A-L1. On average, satellite stellar masses decrease by roughly one order of magnitude, with smaller reductions for the most massive systems. Notably, some satellites are stripped so severely that their final stellar masses fall below  $1\, M_\odot$ (the lower limit of the mass axis of Fig.~\ref{fig:lumfunraddist}). (This explains why the colored lines do not reach the same total $N$ as the black line.) Out of the total $320$ satellites,  a fraction of 25\% (79 satellites) end up with $M_*<1$ M$_\odot$.

Interestingly, the differences  between the Scaled and Fiducial model results are minor.  This result is not trivial: while both models assume the same initial $M_{*,\rm peak}$ stellar masses,  different assumptions about their sizes could, in principle, have led to rather different stripping outcomes. Likewise, the choice of host potential (DMO vs.  Disc) makes little difference, as anticipated from Fig.~\ref{fig:massloss}, where only a few  satellites showed significant differences in the amount of tidally stripped mass.

The total stellar mass lost to tides in Aq-A-L1, and therefore predicted to contribute to its stellar halo, is $2.5\times10^8\,{\rm M}_\odot$. This is lower than current estimates for the stellar halo of the Milky Way \citep[$10^9$ M$_\odot$;][]{Deason2019}. This difference is, however, not unexpected, as the stellar halo mass is dominated by a small number of massive accretion events and our estimate does not include satellites which may have merged with the Milky Way galaxy, such as the Gaia-Sausage-Enceladus (GSE) event \citep[][]{Meza2005,Belokurov2018,Helmi2018}.  In Aq-A-L1, the five most massive satellites already account for $\sim 72\%$ of the total stripped stellar mass, highlighting the strong sensitivity of this quantity to stochastic variations. The presence or absence of even a single massive contributor can have a strong impact on the  total stripped stellar mass.

The right-hand panel of Fig.~\ref{fig:lumfunraddist} shows the radial distribution of Aq-A-L1 satellites. Different linestyles distinguish systems with stellar masses above $100\, M_\odot$ (solid) and above $1000\, M_\odot$ (dot-dashed). Black curves show the $z=0$ radial distribution of the full population of satellites before stripping; i.e., using the   stellar masses at infall time, $M_{*,\rm peak}$.   Since GALFORM predicts a minimum  $M_{*,\rm peak}$ of $\sim 500$ M$_\odot$, the solid line for $M_*>100$ M$_\odot$ is identical to the initial mass distribution of the complete satellite population of Aq-A-L1.

Note that the initial satellite radial distribution is extremely concentrated, with $\sim 40$ satellites predicted to be within $10$ kpc and some  reaching distances as close as $\sim 1$ kpc from the host centre. Very few of these innermost satellites, however, would be detectable because they would be subject to extreme tidal forces. This is shown by the red curves in the right panel of Fig.~\ref{fig:lumfunraddist}  which correspond to the Fiducial model at $z=0$, i.e., after stripping.   For clarity, the equivalent curves for the Scaled model are not shown, since satellite radial distances are assumed identical  in both models, and, as demonstrated in the left panel, the final $z=0$ stellar masses are very similar. Lighter and darker red colors correspond to the DMO and Disc host potentials, respectively. Again, the addition of a disc seems to make little difference to the predicted satellite population, after considering orphans and stripping.

Fig.~\ref{fig:lumfunraddist}  shows that the radial distribution of satellites is strongly affected by stripping, with
essentially no satellites with $M_*>100$ M$_\odot$ remaining  in the innermost $10$ kpc of the host.  This lack of bright systems in the inner regions of the halo is consistent with Fig.~\ref{fig:massloss}, which showed that Aq-A-L1 satellites  with $r_{\rm peri}\lesssim 15$ kpc suffer extreme mass losses. Differences between results obtained using the DMO or Disc potentials are minimal,  although, as expected, the Disc case results in somewhat larger mass losses and, therefore, a slightly less concentrated radial distribution.

These results highlight the significant impact of stellar stripping on both the luminosity function and radial distribution of satellites orbiting within the tidal field of MW-mass halos.  In the next sections we shall use these results to make predictions applicable to the actual MW galaxy. In what follows, we shall only show results corresponding to the Disc host potential but we remind the reader that our results are insensitive to this choice.

\begin{figure}
\centering
\includegraphics[width=\linewidth]{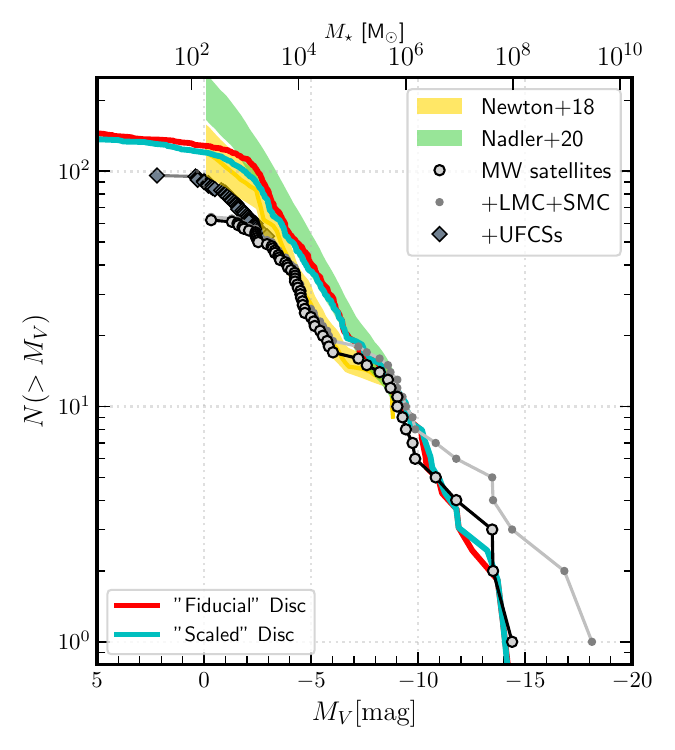}
\caption{
The predicted $z=0$ cumulative $V$-band luminosity function of MW satellites. The red (cyan) line shows the prediction assuming the "Fiducial" ("Scaled") luminosity-size prescription for our stellar stripping model.  We assume the "Disc" host potential. Absolute magnitudes $M_V$ have been estimated from stellar masses assuming a fixed mass-to-light ratio of $M_*/L_V=2$. Grey lines with symbols show currently known observational data for comparison (see App.~\ref{app:data}).  Small grey circles show the luminosity function of all confirmed MW satellite dwarf galaxies. Larger circles show the subset of satellite galaxies excluding the LMC and SMC.
Ultra-faint objects classified as "UFCSs" (i.e., not spectroscopically-confirmed as dwarf galaxies or globular clusters), are included as grey diamond symbols. Shaded bands in the background show predictions from \citet{Newton2018} (yellow) and \citet{Nadler2020} (green), estimated from incompleteness corrections to current surveys.
  }
 \label{fig:MWlumfun}
\end{figure}

\begin{figure*}
\centering
\makebox[\textwidth][c]{
\includegraphics[width=1.1\linewidth]{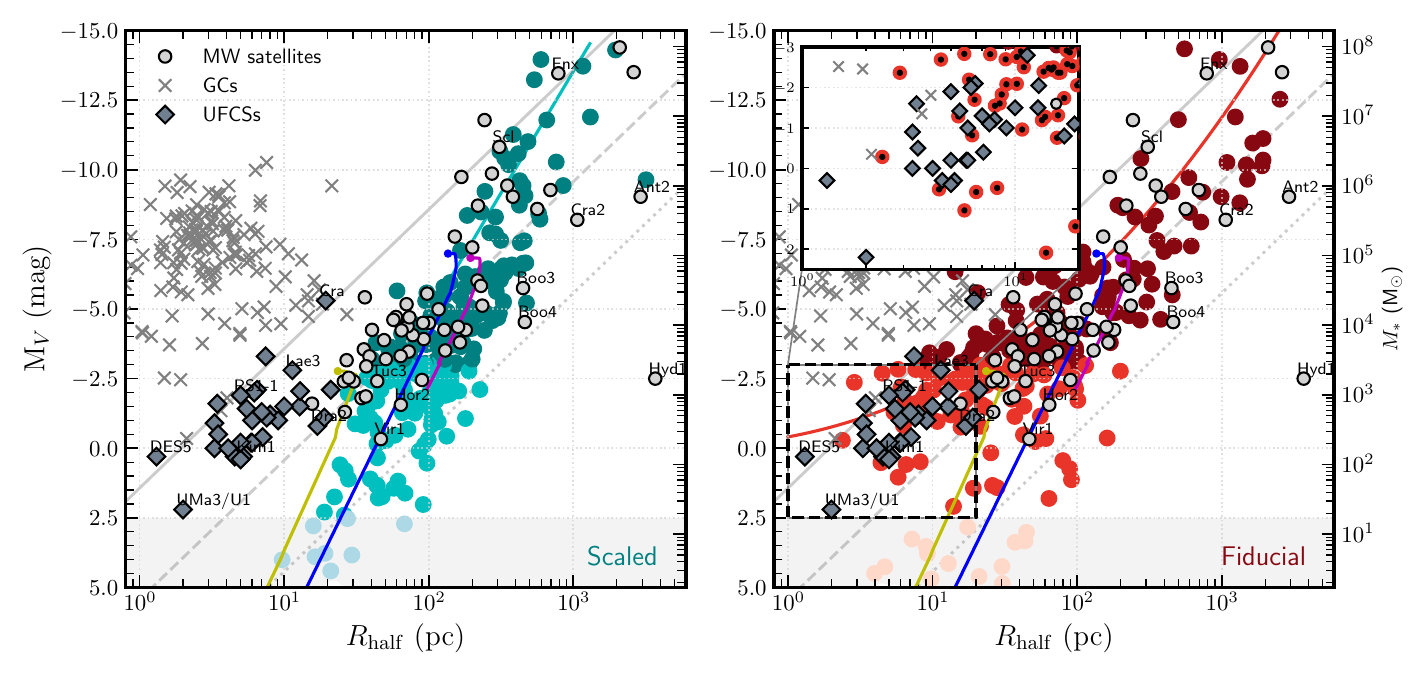}
}
\caption{  
Predicted $z=0$ size-luminosity relation for MW satellites.  
Size  is expressed as the projected half-light stellar radii of satellite galaxies, $R_{\rm half}$.
The left (right) panel shows predictions from our stellar stripping model assuming the "Scaled" ("Fiducial") initial luminosity-size prescription.  We adopt the "Disc" host potential.
Points are colored with different shades according to their $M_V$.
Observational data is overplotted as grey symbols. Circles show confirmed MW satellites,  diamonds show 
objects classified as "UFCSs",  and crosses show  globular clusters (see App.~\ref{app:data}).  A few relevant objects are labelled.
Colored thin lines show the evolution of a few  example satellites.  Some examples end up with very low luminosities, outside of the range plotted. 
For reference, diagonal grey lines in the background represent regions of constant effective surface brightness of $(25, 29, 33)$ mag/as$^2$, shown in solid, dashed, and dotted linestyles, respectively.
A grey shade covers the luminosities ($M_V>5$) below which systems would become undetectable for practical purposes. In the right panel,  the box identifies model satellites that overlap with observed UFCSs, i.e.,  satellites with $2.5>M_V>-3$ and $1<R_{\rm half}<20$ pc.
The inset panel zooms into this region to enable a more detailed comparison between model and data.
  }
 \label{fig:MWMVrhalf}
\end{figure*}

\subsection{Predicted MW satellite luminosity function}\label{sec:MWsatlum}

Figure~\ref{fig:MWlumfun} shows, with solid lines, the predicted cumulative luminosity function of MW satellites at $z=0$ after stripping.  
Assuming satellite numbers scale linearly with host halo mass \citep{Springel2008},
MW predictions are computed by rescaling the  total number of Aq-A-L1 satellites  down by a factor of $1.6$ to take into account that the Aq-A virial mass ($1.8\times 10^{12}$ M$_\odot$) is higher than the $(1.1\pm 0.2)\times 10^{12}$ M$_\odot$ inferred for the MW  \citep[e.g.][]{Cautun2020}. Because of this scaling, our model implies a total number  of $196\pm 41$ MW satellites of all luminosities.  The associated uncertainty comes from propagating the $20\%$ uncertainty in the MW's virial mass, plus the intrinsic statistical uncertainty associated with the CDM subhalo mass function, expected to follow Poisson statistics  \citep{Wang2012}.
The combination of these independent contributions in quadrature yields a total uncertainty of $\sim 21\%$.  (We exclude here the uncertainty associated with the assumed redshift of reionisation, as $z_{\rm rei}=6$ is relatively well-constrained.)

Luminosities in Fig.~\ref{fig:MWlumfun} are expressed as absolute magnitudes in the V-band, $M_V$, computed from the model stellar masses assuming a constant mass-to-light ratio of $M_*/L_V=2$ in solar units \citep{Woo2008}. The predicted number of MW satellites is shown for both the Scaled (cyan) and Fiducial (red) galaxy-size prescriptions assuming the Disc host potential.

As expected from Fig.~\ref{fig:lumfunraddist}, the Scaled and Fiducial models yield very similar $z=0$ MW satellite luminosity functions after stripping. In both cases, the number of satellites grows steadily with decreasing luminosity down to  $M_V\approx -2$,  beyond which the number stagnates and only increases slowly at fainter magnitudes.

Grey symbols show the current observed inventory of MW satellites.  Small filled circles correspond to all confirmed MW satellites, while larger circles denote the same systems, but excluding  the Magellanic Clouds (LMC and SMC). The Aq-A-L1 simulation does not host LMC- or SMC-mass satellites at $z=0$, which is perhaps not suprising as these are uncommonly luminous satellites around MW-mass hosts \citep[e.g.,][for a recent discussion]{Santos-Santos2021}.

Note that, after removing the Clouds, the predicted satellite luminosity function matches the MW one quite well, especially in the classical satellite regime, $M_V<-8$.  At fainter luminosities ($M_V>-8$) our predictions follow the data closely but show a small offset, which increases towards fainter magnitudes. Our scaled Aq-A-L1 model predicts a total of roughly $82 (41)$ satellites brighter than $M_V=-3 (-5)$, whereas only $51 (26)$ such systems are known around the MW. 

\subsubsection{UFCS candidates}

Darker grey diamonds in Fig.~\ref{fig:MWlumfun} show the MW satellite function after adding the UFCS MW satellite candidates.  These compact systems are rather faint, with luminosities roughly between  $2.5>M_V>-3$. This magnitude interval coincides  with the range of luminosities where our model predicts a substantial excess of satellites over the current census.  Specifically, the model predicts $\sim 132$ satellites brighter than $M_V\sim 1$. It is therefore tempting to associate at least some of these UFCS candidates with the dwarf galaxy satellites predicted by our model.  

This hypothesis is supported by the results of other studies, which have also tried to estimate the abundance of ultra-faint MW satellites, using mainly empirical models to correct for incompleteness in our current census. 
Specifically, the yellow and green bands in Fig.~\ref{fig:MWlumfun} correspond to the MW satellite luminosity function from \citet{Newton2018} and \citet{Nadler2019,Nadler2020}\footnote{The more recent \citet{Tan2025} satellite count estimate is consistent with \citet{Nadler2020}.}, respectively ($\pm 1\sigma$ bands).

Our predictions lie somewhere in between  these two studies, although one should perhaps not overinterpret this agreement, as those studies adopt quite different prescriptions from ours for modelling key ingredients such as orphan galaxies and  the radial distribution of satellites. In addition, these previous studies did not consider that the missing ultra-faint satellites could be as compact as the  UFCS population.

Another important difference is that the Aq-A halo does not have an LMC-mass satellite, which may have contributed its own satellites to the MW \citep[e.g.][]{Santos-Santos2021}. Because of these complications, we refrain from a detailed comparison with either study, but note that the prediction of a substantial extra population of ultra-faint MW satellites appears to be a robust prediction of $\Lambda$CDM models.

We end by noting that, although our predictions are based on a single halo (Aq-A-L1), halo-to-halo scatter in  the total number of satellites is expected to be dominated by Poisson uncertainties 
(of order $\sim 7\%$ for a total of $\sim196$ satellites),
with limited effect on our estimates at the low-mass end of the satellite luminosity function.  In this regime, the abundance of ultra-faint systems is largely set by the overall halo mass and is therefore already captured by the $\sim 21\%$ uncertainty quoted in the previous subsection.  Halo-to-halo variations are expected to become more important only at the high-mass end, where stochasticity in the accretion history can lead to significant differences, as illustrated, for example, by the presence of the Magellanic Clouds in the Milky Way compared to Aq-A.

This picture is supported by the work of \citet{Wang2012},  who show that in dark-matter-only simulations different assembly histories do not introduce additional scatter in the subhalo abundance beyond Poisson noise at fixed halo mass. It remains unclear, however, whether this conclusion carries over once luminous satellites  are considered and if particular accretion histories, such as an early massive merger analogous to the Gaia-Enceladus-Sausage,  could  leave an imprint on the satellite luminosity function beyond what is captured by our uncertainty estimate, even if their effect on the underlying subhalo population might be small. We therefore caution that the $\sim 21\%$ figure is best regarded as a rough estimate rather than a definitive error budget, and that a proper quantification of halo-to-halo scatter will require applying our framework to a suite of MW-mass halos, a task we defer to future work.

\subsection{Predicted size-luminosity relation of MW satellites}\label{sec:MWsatsize}

The colored symbols in Fig.~\ref{fig:MWMVrhalf} show the predicted $z=0$ size-luminosity relation for MW satellites.  The left panel presents results assuming the Scaled model for galaxy sizes, which, to recall, makes the simple assumption that initial galaxy size is simply proportional to the characteristic radius of their dark matter subhalo at infall. As a result, most satellites have initial sizes which exceed a few tens of parsecs, and are therefore much larger than the MW UFCSs. On the other hand, the right-hand panel shows results for the Fiducial model, which allows for the formation of ultra-compact ``microgalaxies'' (comparable in size to UFCSs) at the very faint end of the luminosity function.

Different shades of cyan (left panel) or red (right panel)  distinguishes satellites of different luminosity,  split into three categories:  $M_V<-3$ (darker shade), $2.5>M_V>-3$ (intermediate),  and $M_V>2.5$ (lighter). 
Grey symbols correspond to MW observational data: confirmed satellite galaxies are shown as circles, UFCSs as diamonds, and globular clusters (GCs) as crosses. A few select objects are labelled, for reference. Diagonal grey lines in the background identify lines of constant effective $V$-band surface brightness of $(25, 29, 33)$ mag/arcsec$^2$, shown by solid, dashed, and dotted line styles, respectively.

The location of MW satellite galaxies in the $M_V$-$R_{\rm half}$ plane at $z=0$ predicted by the Scaled and Fiducial models differ substantially.  In the Scaled model (left panel of Fig.~\ref{fig:MWMVrhalf}), heavily stripped satellites ($M_V>-3$) extend to fainter magnitudes the trend defined by the initial $M_{*,\rm peak}$–$R_{\rm half,peak}$ relation (indicated by a colored solid line in the background, see Fig.~\ref{fig:MstRh}).

For reference,  evolutionary tracks for a few illustrative examples are overdrawn as thin coloured lines, with small circles marking the initial position of each satellite in the $M_V$–$R_{\rm half}$ plane.  The tracks are steep,  showing that tides mainly act to reduce a satellite's luminosity, without changing much its size \citep[][]{Errani2024b}. Individual tracks are roughly parallel to each other, and in a couple of the cases evolve to luminosities fainter than the bottom limit of the plot,which corresponds to roughly $M_*\sim 1$ M$_\odot$.

Because in the Scaled model the faintest galaxies at infall ($M_* \sim 10^3$  M$_\odot$, or $M_V\sim -2$) had initial sizes of $50$-$100$ pc, 
we find no "detectable" satellites (i.e., $M_V<2.5$, or $M_* \gtrsim 17\,  M_\odot$) with sizes smaller than roughly  $10$ pc. This implies that, in this model, the UFCS population, with half-light radii between $1$-$20$ pc, cannot be the tidal remnants of luminous dwarfs.

In other words, if UFCSs are indeed dwarf galaxies, then they must have been born more or less as compact as they are found at present. This is actually one of the main motivations for introducing the Fiducial model, which allows for a population of initially compact systems at the faint end (see Fig.~\ref{fig:MstRh}).  Indeed,  the Fiducial model  (right-hand panel of Fig.~\ref{fig:MWMVrhalf}) yields a final distribution of sizes that overlaps the UFCS population.
Systems as compact as those inhabit the dense cusp of their cold dark matter halos and are thus hardly perturbed by tidal evolution, remaining more or less close to their initial positions in the mass-size plane.

Extreme stripping of more luminous (and larger) systems can still lead to the formation of extremely faint objects in the Fiducial system too, but those tidal remnants are still expected to be relatively large as their final size would reflect the initial size of their progenitors (see the faintest systems in the right-hand panel of Fig.~\ref{fig:MWMVrhalf}).

Finally, we note that because the Scaled and Fiducial models represent two physically motivated extremes (initial galaxy sizes that scale with the dark halo characteristic radii, or a stellar mass-size relation that allows for the presence of initially compact ultra-faint dwarf galaxies), any intermediate assumption for the initial size-luminosity relation would yield an intermediate degree of overlap with the UFCS population. In what follows, we shall focus on the Fiducial model, which maximises this overlap, to explore the properties of model satellites in the UFCS regime and to gain some insight into their predicted orbits, radial distribution, and internal kinematics.

\subsection{Predicted radial distribution of MW satellites}\label{sec:MWsatdist}
\begin{figure}
\centering
\includegraphics[width=\linewidth]{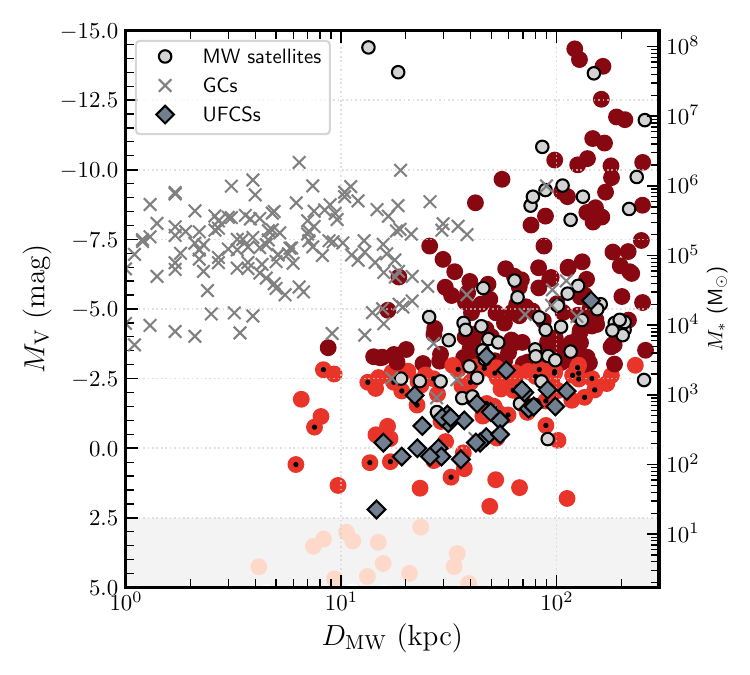}
\caption{
Predicted $z=0$ luminosity vs Galactocentric distance $D_{\rm MW}$ for MW satellites.  The "Fiducial" luminosity-size prescription and Disc potential is used for our stellar stripping model.  As in previous figures, grey symbols correspond to observational data.  A grey shade indicates the luminosities ($M_V>5$) below which systems would become undetectable for practical purposes. Small black dots highlight the model galaxies that overlap observed UFCS systems in luminosity and size (see box in Fig.~\ref{fig:MWMVrhalf}).
}
 \label{fig:MWdist}
\end{figure}

Figure~\ref{fig:MWdist} shows the predicted radial distribution of satellites, shown as a scatter plot of absolute magnitude, $M_V$, as a function of Galactocentric distance,  $D_{\rm MW}$, at $z=0$. As in Fig.~\ref{fig:MWMVrhalf}, predictions of the Fiducial model are shown as coloured circles, where colour-coding indicates absolute $V$-band magnitude (lighter shades correspond to fainter luminosities). Grey symbols represent observational data. In order to directly compare Aq-A-L1 results with MW data, we rescale the model Galactocentric distances by the ratio of  Aq-A-L1's virial radius size ($r_{200}=245.7$ kpc) and that expected for the MW. (We assume $r_{\rm 200,MW}=213.4$ kpc, which corresponds to the virial radius of an NFW halo with $M_{200}=1.1\times 10^{12}$ M$_\odot$ and average concentration, following \citealt{Cautun2020}.) Small black dots indicate the model galaxies that overlap with observed UFCS systems in luminosity and size (see box in Fig.~\ref{fig:MWMVrhalf})

Fig.~\ref{fig:MWdist} illustrates an interesting result: essentially all ``detectable'' satellites, defined as those with $z=0$ luminosities brighter than $M_V\sim 2.5$ ($M_*\gtrsim 17\, M_\odot$ for our assumed mass-to-light ratio of 2 in solar units), are predicted to lie at distances larger than $\sim10$-$15$ kpc away from the MW centre. This distance ``threshold'' is consistent with the discussion of Fig.~\ref{fig:massloss}, where we showed that  satellites with pericentric distances smaller than $\sim 15$ kpc suffer extreme tidal stripping losses.

Interestingly, our predictions agree remarkably well with the distribution of Galactocentric distances observed for both confirmed MW satellites and the UFCS population, all of which have $D_{\rm MW}>10$ kpc.  In addition, the majority of globular clusters are found much closer to the Galactic centre, typically within $10$-$20$ kpc. This supports the interpretation that at least some of the UFCSs are not star clusters, but actually members of the dwarf galaxy satellite population. 

\subsection{Predicted mean densities of MW satellites}\label{sec:MWsatdens}
\begin{figure*}
\centering
\includegraphics[width=0.49\linewidth]{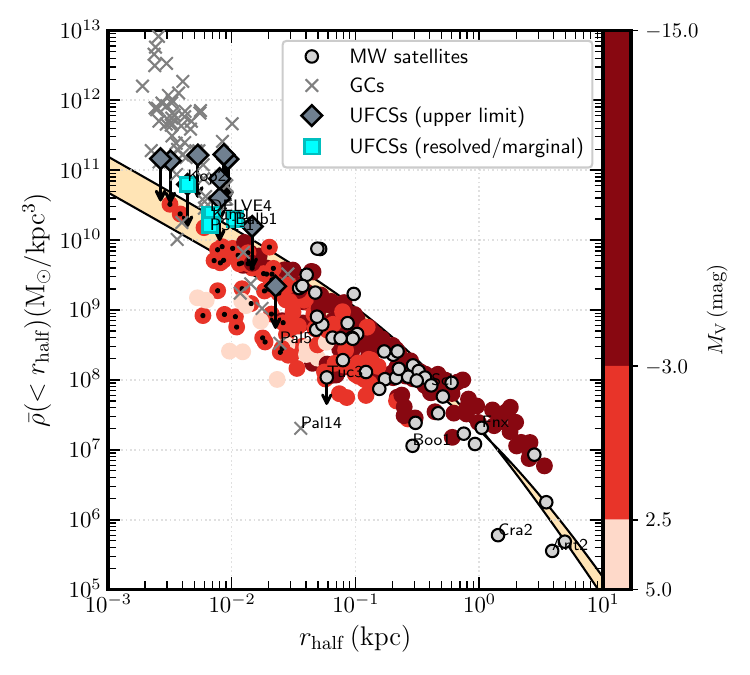}
\includegraphics[width=0.49\linewidth]{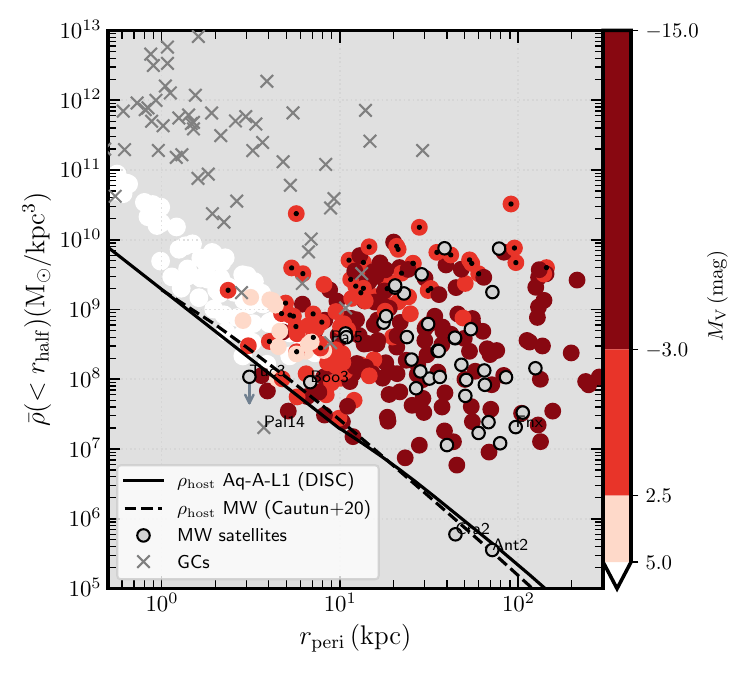}
\caption{ 
\textit{Left: } Predicted $z=0$ mean  density within the half-mass radius, $\bar{\rho}(<r_{\rm half})$ vs $r_{\rm half}$ for MW satellites.  As in previous figures, predictions from Aq-A-L1 are shown in shades of  red while observational data are shown in in grey. Densities for Aq-A-L1 satellites are computed as described in Sec.~\ref{sec:MWsatdens}. 
The "Fiducial"  luminosity-size model with Disc potential  is assumed to estimate stellar stripping. 
For observational data, densities are estimated from velocity dispersion  ($\sigma$) and half-light radii measurements using \citet{Wolf2010}'s mass estimator. 
The subset of UFCS systems with available $\sigma$ data constraints  are shown as
cyan squares  with labels (if $\sigma$ is resolved or marginally-resolved), or as dark grey diamonds (if only upper limits are available; see App.~\ref{app:data}).
Small black dots mark the model galaxies that overlap with observed UFCS systems in luminosity and size (see box in Fig.~\ref{fig:MWMVrhalf}).
\textit{Right: }Predicted  mean density vs pericentric distance $r_{\rm peri}$, for MW satellites.  The solid line shows the density profile of the host Aq-A halo assuming the  Disc potential. The dashed line shows the corresponding profile estimated for the Milky Way Galaxy according to \citet{Cautun2020}. White points indicate predicted satellites with $z=0$ luminosities fainter than $M_V>5$ ($\approx M_*<1$ M$_\odot$).  Small black dots mark the model galaxies that overlap with observed UFCS systems in luminosity and size (see box in Fig.~\ref{fig:MWMVrhalf}).
  }
 \label{fig:MWrho}
\end{figure*}

Another feature that may be used to distinguish GCs from dwarf galaxy satellites is the mean density inside their 3D half-light radius, estimated as $r_{\rm half}=(4/3)R_{\rm half}$. 
We compute mean densities for Aq-A-L1 satellites at $z=0$ using the truncated NFW profile described in \citet{Errani2021} (see their eqs. 7-9).
At any given time, this profile  is fully specified by $r_{\rm max}$, $V_{\rm max}$, and the remaining mass bound fraction, $M_{\rm max}/M_{\rm peak}$. We integrate to obtain the spherically-averaged mean enclosed density profile and interpolate to determine its value at $R_{\rm half}$\footnote{A description of the evolution of $\bar{\rho}(r_{\rm half})$ and $r_{\rm half}$ for Aq-A-L1 satellites, from infall time to $z=0$, is given in App.~\ref{app:dens} and Fig.~\ref{figapp:rhorhalf}.}.

For observed satellites, we estimate densities within the 3D half-light radius from the  line-of-sight velocity dispersion $\sigma$ and $R_{\rm half}$ measurements using the dynamical mass estimator of \citet{Wolf2010},
$M(<r_{\rm half}) = 4\sigma^2\, R_{\rm half} / G $
where $M(<r)$ is the enclosed total mass profile. 
With this assumption,  and dividing by the volume of a sphere of radius $r_{\rm half}$, 
\begin{equation}\label{eq:dens}
\bar{\rho}(<r_{\rm half})= 1.27 \sigma^2 / (G \pi R_{\rm half}^2).
\end{equation}

The left panel of Fig.~\ref{fig:MWrho} shows $\bar{\rho}(<r_{\rm half})$ vs $r_{\rm half}$. 
The yellow band in the left panel of Fig.~\ref{fig:MWrho} shows the size-density relation predicted at infall for $\Lambda$CDM halos with mass comparable to the critical mass for galaxy formation, $M_{\rm crit}$, discussed in Sec.~\ref{SecCritMass}. 
Specifically, these lines trace NFW profiles of  $V_{\rm peak}=20$ km/s and average concentration at $z_{\rm peak}=3$ and $15$,  which correspond to halos of masses $M_{200}\approx 5\times 10^8$ and $7\times10^7$ M$_\odot$.

GCs are extremely dense, with values typically exceeding $10^{10}\, M_\odot$/kpc$^3$ and reaching in some cases $10^{12}\, M_\odot/$kpc$^3$. Observed dwarf galaxies (grey circles), on the other hand, have much lower densities, spanning the range $10^{7}$ to $10^{10}$ M$_\odot/$kpc$^3$. There is also a clear trend for smaller dwarfs  to be denser, as expected if they inhabited a cuspy dark matter halo.

Tides are expected to remove preferentially dark matter, and, therefore, to shift satellites to the region below the yellow NFW band, as shown by the faint dwarf galaxies in the model (red circles), which overlap many of the grey circles corresponding to observed MW dwarfs. Note that  this agreement is not guaranteed by our assumptions, which only set the initial size of galaxies as a function of stellar mass. We discuss the effect of tides in this plane in more detail in  Appendix~\ref{app:dens}.

On the other hand, dwarf galaxies largely unaffected by tides remain close to the yellow band. Such galaxies should have densities that increase with decreasing $r_{\rm half}$, reaching $\sim 10^{10}$ M$_\odot/$kpc$^3$ for galaxies with $r_{\rm half}$ of the order of $10$ pc, and $10^{11}$ M$_\odot/$kpc$^3$ for those with $r_{\rm half}\sim 1$ pc. This dependence on size of the mean density is particularly important for compact systems such as UFCSs, whose half-light radii probe exactly this regime.

\subsubsection{Comparison with UFCSs}

Recently,  velocity dispersion measurements have become available for a subset of UFCSs, enabling density estimates for some of these systems \citep[][]{Cerny2026}. At present, five dwarfs have preliminary velocity dispersion determinations (Balbinot 1,  DELVE 4, Kim 1, PS1-1, Koposov 2)\footnote{Note that the datapoints for DELVE 4 and Kim 1 are on top of each other, as they present the same densities and half-light radii.}, and their location on the size-density plane is shown by  
large cyan squares.
 Interestingly, these UFCSs follow the trends highlighted above: they appear to be denser than other (larger) UFDs, and lie close to the edge of the yellow NFW band expected for dwarfs unaffected by tides. This supports our interpretation that UFCSs are so dense (because they sit deep inside the dark matter cusp) that they are very resilient to tides. Other 
systems
  with kinematic data (see grey diamonds with downward arrows in the left panel of  Fig.~\ref{fig:MWrho}) have only upper limits to their velocity dispersions \citep{Cerny2026}.
While these upper limits cannot be taken as positive evidence for high densities, their values are at present consistent with this scenario.

The right panel of Fig.~\ref{fig:MWrho} shows again mean satellite density, but now as a function of pericentric distance. The black lines in this figure show the mean enclosed density profile of the Aq-A-L1 host (solid line, assuming the “Disc” potential) and of the Milky Way (dashed line; \citealt{Cautun2020}). Both profiles are very similar; in addition, as expected, most observed dwarfs and GCs are typically denser than the host halo at their pericentre\footnote{Grey points and crosses show observed satellite galaxies and globular
  clusters, respectively, for which both pericentric distances and velocity dispersion data are available.  Robust orbital parameter estimations are not yet available for UFCSs (see App. \ref{app:data} for details).}.
Only a few exceptions exist, such as Tuc 3,  Cra 2 and Ant 2, which appear to be in the process of being tidally disrupted  \citep[see e.g.][]{Drlica-Wagner2015,Shipp2018,Ji2021,Pace2022}.

Model dwarf galaxies (coloured points) overlap observed dwarf galaxy satellites not only  in density but also  in pericentric distance. As seen in Fig.~\ref{fig:massloss}, satellites with $r_{\rm peri}\lesssim 15$ kpc experience extremely strong tidal stripping, as this distance roughly marks where the subhalo peak density equals the host's mean enclosed density. Consequently,  most satellites with $r_{\rm peri}\lesssim 15$ kpc are stripped to stellar masses $<1$ M$_\odot$, and thus  rendered  undetectable by $z=0$ (white circles in right panel of Fig.~\ref{fig:MWrho}). Our modeling thus predicts that detectable dwarf galaxy satellites should have larger pericentric distances than GCs and at least an order of magnitude lower central densities.

In summary, our results show that the ultra-faint satellite galaxies predicted by our study should have densities and Galactocentric/pericentric distances that are clearly distinct from those of globular clusters. This makes strong predictions for the mean density of the observed UFCS population if they are, in fact, dwarf galaxies. We explore predictions for the velocity dispersions of UFCSs next.

\begin{figure*}
\centering
\includegraphics[width=0.7\linewidth]{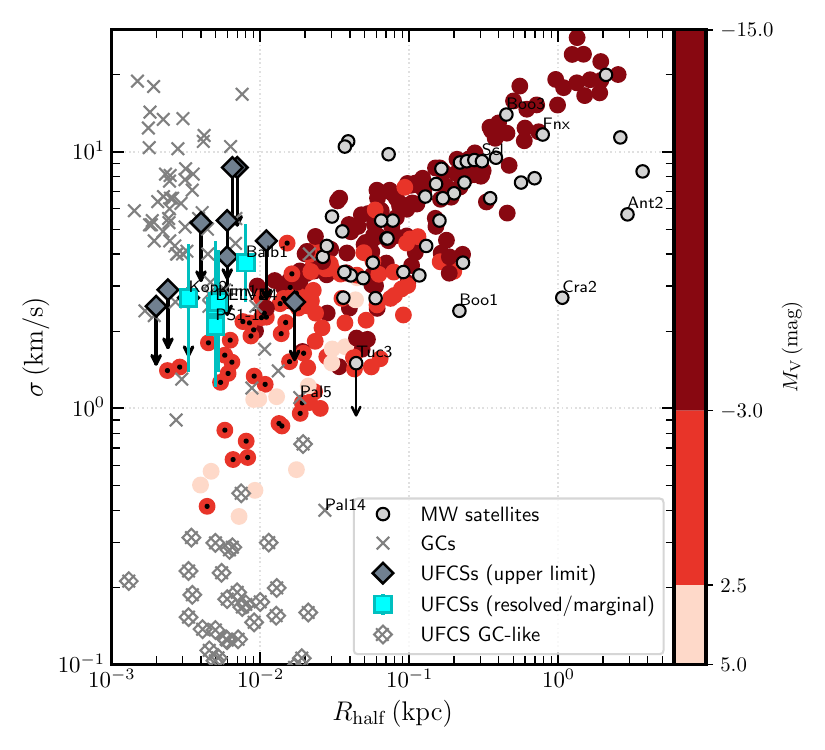}
\caption{ 
Predicted $z=0$ velocity dispersion ($\sigma$) vs projected half-light  radius ($R_{\rm half}$) for MW satellites.
The Fiducial Disc luminosity-size prescription  is assumed.
As in previous figures, predictions from Aq-A-L1 are shown in red and observational data in grey (see App.~\ref{app:data}).
Small black dots mark the model galaxies that overlap with observed UFCS systems in luminosity and size (see box in Fig.~\ref{fig:MWMVrhalf}).
The subset of UFCS systems with available $\sigma$ data constraints  are shown as cyan squares with one standard deviation uncertainties and labels (if resolved or marginally-resolved) or as dark grey diamonds (if upper limits; see App.~\ref{app:data}). 
We also estimate the velocity dispersion that UFCS objects would have if they were globular clusters and represent them as open diamond symbols with a cross.
  }
 \label{fig:MWsigma}
\end{figure*}

\subsection{Predicted velocity dispersions of MW satellites}\label{sec:MWsatsigma}
We estimate the velocity dispersion of stars in our simulated satellites 
using the dynamical mass estimator of \citet{Wolf2010},
$\sigma\approx \sqrt{(G M(<r_{\rm half}) ) / (3 r_{\rm half})}$, where $M(<r_{\rm half})$ is the total mass enclosed within the 3D half-light radius computed from the truncated NFW density profile described in Sec.~\ref{sec:MWsatdens}.

Figure~\ref{fig:MWsigma} shows the predicted $\sigma$ for satellites in our model as a function of their projected half-light radius, $R_{\rm half}$. Points are colored according to their $z=0$ luminosities.
As in Figs.~\ref{fig:MWdist} and \ref{fig:MWrho}, the small black dots in Fig. ~\ref{fig:MWsigma}  identify the subset of model satellites that share the same luminosity and size as the observed UFCS population (i.e., which fall inside the inset box in Fig. ~\ref{fig:MWMVrhalf}).
Data for observed systems are overplotted in grey, with UFCSs 
as cyan squares with uncertainties (if resolved or marginally resolved) or dark grey diamonds with downward arrows (if upper limits only; see App. ~\ref{app:data}).
As discussed above, only five UFCS systems (Balbinot 1, PS1-1, DELVE 4, Kim 1, Koposov 2) have reasonably firm velocity dispersion estimates (squares), while the rest are upper limits due to their extremely faint luminosities and small numbers of member stars bright enough for spectroscopic follow-up \citep{Cerny2026}.  

Our predictions for satellites brighter than $M_V< -3$ are in very good agreement with the data for confirmed MW satellites, a result that implies that the dark matter content of our model galaxies is in good agreement with that of observed dwarfs. Note that this is not guaranteed by our assumptions, and provides strong support for the galaxy mass-halo mass relation in our model, and, indirectly, for the cuspy nature of the inner density profiles of dark matter subhalos.

For ultra-faint satellite galaxies our results make two main predictions: that velocity dispersions should decrease as the size of the ultra-faint decreases, and that UFCSs, if truly dwarf galaxies, should have $\sigma\sim 1$-$3$ km/s.

For comparison, we compute  as well an estimate of the velocity dispersions that the UFCS population would have if they were globular clusters, i.e.,  self-gravitating stellar systems with no dark matter. To compute the latter, we follow eq.2  in  \citet{Errani2024}, i.e.,
$\sigma^2_{\rm sg}\approx (5/96) \, G  \,  M_* / r_*$,
where the scale radius $r_*$ is related to the projected half-light radius by $R_{\rm half} \approx  2.02\, r_*$. These results are shown in Fig.~\ref{fig:MWsigma} as open diamonds with crosses, and indicate that, if UFCSs are star clusters with no dark matter, then their velocity dispersions should be of order $100$-$300$ {\it meters} per second. Such low values of $\sigma$ are likely beyond current observational capabilities at present.

Our results emphasize that precise measurements of velocity dispersions offer the most robust means of distinguishing the true nature of the UFCS population. Unfortunately, such low velocity dispersions remain extremely challenging to measure with current instrumentation. Even with Keck/DEIMOS \citep[e.g.,][]{Cerny2025}, which combines relatively high spectral resolution and large collecting area, the attainable precision is only of the order of a few km/s for satellites as faint and distant as UFCSs.

Additional uncertainties arise from the small number of candidate member stars and possible contamination from unresolved binaries.
One example is UMa3/UNIONS1 \citep[the faintest MW satellite known to date, with $M_V\sim 2.2$;][]{Smith2024,Errani2024,Devlin2025}, which has roughly a dozen member stars within spectroscopic reach. The latest available data is unable to distinguish between the GC or dwarf hypothesis; even after two epochs of observation one can only set upper limits to the velocity dispersion ($\sigma<2.5$ km/s at $95\%$ confidence), which is consistent with either interpretation \citep{Cerny2025,Cerny2026}.

\section{Summary and Conclusions}\label{sec:conclu}

We have presented a detailed model of the abundance, radial distribution, orbital parameters, and internal properties of the \textit{full} population of present-day satellites around Milky Way (MW)-mass halos in $\Lambda$CDM. This model is based on  Aq-A-L1, the highest-resolution  dark matter-only N-body simulation of a MW-mass halo available at present, combined with the GALFORM semianalytical galaxy formation model. Our model assumes that galaxies only form in subhalos that exceed the hydrogen-cooling ``critical mass'' threshold of \citet{Benitez-Llambay2020}, all of which are well resolved in Aq-A-L1.

The critical mass assumption implies that only subhalos with peak circular velocities $V_{\rm peak}\gtrsim 15$ km/s  can host luminous galaxies. GALFORM also predicts that the stellar mass of subhalos should scale like  a steep power-law of $V_{\rm peak}$. The model further tracks every luminous subhalo ever accreted into the Aq-A-L1 halo and follows its dynamical evolution until the present time, including both ``surviving'' subhalos and ``orphan'' systems whose subhalos have artificially disrupted due to limited resolution.

For the first time in models of this kind, we extend this framework to include an semi-analytic treatment of both dark matter and stellar tidal stripping, enabling robust predictions for the MW satellite population at $z=0$.  Tidal effects are computed using the tidal track formalism of \citet{Errani2021,Errani2022}, who argue that cuspy CDM subhalos generally survive tidal mass losses.

The tidal tracks follow the mass of stars and dark matter that remain bound to a satellite as a function of time. Tidal effects in this model depend critically on the host potential, the pericentric distance of a satellite, the radial segregation of the stellar component within its dark matter halo, and the number of orbits completed since first infall.

Our model considers two host potentials to compute tidal effects: the DMO Aq-A-L1 potential at $z=0$, and a “Disc’’ case  where the presence of a baryonic disc is assumed to flatten the inner circular velocity curve of the galaxy inside $10$ kpc from the centre, roughly consistent with current constraints for the MW.

Our results are especially sensitive to the size of the stellar component. We find that GALFORM predicts dwarf galaxy sizes well in excess of those of observed dwarfs, so we adopt two simple models for the size-luminosity relation of satellite galaxies at infall: a ``Scaled'' model where the stellar half-light radii are simply proportional to the characteristic radii of their dark matter halos (this model implicitly sets a lower size limit of roughly $50$-$100$ pc for the faintest satellites), and another, ``Fiducial'', model where compact galaxies with much smaller sizes, down to a few parsecs, are allowed to form at the faint end.

Applied to Aq-A-L1,  our modeling yields results for the MW satellite population that may be summarized as follows.

\begin{itemize}

\item The total number of luminous satellites in Aq-A-L1 cannot exceed the number of subhalos with masses above the ``critical''
hydrogen-cooling
 threshold for galaxy formation assumed in our model.  This corresponds to roughly $320$ in total, all of which are well resolved in the simulation. After scaling to take into account that the virial mass of Aq-A-L1 is roughly $1.6$ times the accepted MW virial mass, our model implies an upper limit of $196\pm 41$ for the total number of satellites of all luminosities around the MW. The quoted uncertainty includes the error in the MW's estimated virial mass as well as Poisson noise coming from the subhalo mass function.  Halo-to-halo scatter is not expected to affect much these estimates but they are sensitive  to the assumed redshift of reionization ($z_{\rm reion}=6$ assumed throughout).

\item Even at the resolution of Aq-A-L1, more than half of the luminous subhalos are ``orphans'' at $z=0$ (i.e., have been artificially disrupted because of numerical limitations, see \citetalias{Santos-Santos2025}), which implies that predictions for the full MW satellite population should include a careful treatment of the expected tidal effects on the stellar and dark matter component of luminous subhalos.

\item As expected, tidal effects are strongest on satellites with small pericentric distances and short orbital times, which affect more severely satellites accreted early into the MW halo. The model predicts that satellites with $r_{\rm peri}\lesssim15$ kpc lose so many stars that they are, for practical purposes, undetectable. 
This effectively reduces the total predicted number of MW satellites from $196\pm 41$ to $138\pm 29$ (with $M_*>10$ M$_\odot$ or $M_V<3$), and implies that very few dwarf galaxy satellites, if any, should be found in the inner $10$-$15$ kpc of the MW.

\item Unlike previous work, we argue that although the addition of a central ``Disc'' to the host potential may increase the number of ``orphan'' satellites in a given N-body simulation, it does not have a large effect on the  final results of the modeling, once tidal effects and orphan satellites are properly taken into account.
(This conclusion applies specifically to our spherical ``Disc" potential, an approximate representation of the deepening of the inner potential due to a baryonic component, and it does not  include the effects of a flattened Galactic disc, with its associated disc shocking and non-spherical torques.)

\item Scaled to the MW, the Aq-A-L1 luminous satellite population matches well the bright end of the MW satellite luminosity function, and predicts that the number of ultra-faint dwarfs should be roughly twice as many as currently known. 

\item
When we assign sufficiently small initial sizes to ultra-faint galaxies,  we predict that
these ``missing'' ultra-faint dwarfs have properties that overlap those of the  UFCSs recently identified in panoramic surveys of the MW halo.
Assuming instead an initial galaxy size-luminosity relation   where  stellar sizes  track the characteristic radius of their dark matter halo, no such overlap between our model and the UFCS population is found (Fig. \ref{fig:MWMVrhalf}). 

\item  
Our results suggest that dwarf-galaxy UFCSs should be at moderate Galactocentric distances ($10$-$150$ kpc), and should be heavily dark matter dominated, with velocity dispersions of order $1$-$3$ km/s which decrease steadily for smaller systems. This is an order of magnitude higher than expected for UFCSs if they were dark matter-free star clusters.

\item Our results also predict that UFCSs should be extremely dense, and that their density, measured inside their half-light radius, should increase towards smaller systems.

\item 
Our results suggest that these UFCSs are not the tidal remnants of once more luminous satellites, but rather ``microgalaxies'' born so deep in the cusp of their dark matter halos that they have been largely unaffected by tidal effects to the present day. 

\end{itemize}

The main conclusions of this study, namely, that the full population of MW satellites does not exceed a few hundred systems, and that many of the UFCSs are dark matter-dominated dwarf galaxies, are eminently falsifiable, and will be put to the test by ongoing and upcoming deep imaging surveys such as
DES, HSC-SSP, DELVE, UNIONS, and LSST, supplemented by spectroscopic follow-up of as many stars as possible in UFCSs.

Confirming that these compact MW satellites are indeed dark matter dominated would be a major result,
and direct proof of the ability of dark matter to cluster to form tide-resilient, high-density cusps at the centres of dark matter halos, a cornerstone prediction of  $\Lambda$CDM.
Furthermore, it would imply that ultra-faint dwarf galaxies are able to form with extremely compact initial sizes, opening a new window on galaxy formation at the very faint end.

\section*{Acknowledgements}
We thank Rapha\"el Errani for guidance in using his tidal track models, and Will Cerny for kindly sharing data for UFCS satellites.  ISS and CSF acknowledge support from the European Research Council (ERC) Advanced Investigator grant to C.S. Frenk, DMIDAS (GA 786910) and from the Science and Technology Facilities Council [ST/P000541/1] and [ST/X001075/1].  The simulations for the Aquarius Project were carried out at the Leibniz Computing Centre, Garching, Germany, at the Computing Centre of the Max-Planck-Society in Garching, at the Institute for Computational Cosmology in Durham, and on the ‘STELLA’ supercomputer of the LOFAR experiment at the University of Groningen.  This work used the DiRAC@Durham facility managed by the Institute for Computational Cosmology on behalf of the STFC DiRAC HPC Facility (www.dirac.ac.uk). The equipment was funded by BEIS capital funding via STFC capital grants ST/K00042X/1, ST/P002293/1, ST/R002371/1 and ST/S002502/1, Durham University and STFC operations grant ST/R000832/1. DiRAC is part of the National e-Infrastructure.  JFN acknowledges the support of the Max-Planck Institute for Astrophysics, of the Donostia International Physics Centre, and of the Physics Department at Durham University.

\section*{Data Availability}
The simulation data underlying this article may be shared upon
reasonable request to the Virgo Consortium's steering committee.

\bibliographystyle{mnras}
\bibliography{archive} 

\appendix
\section{Observational data used}\label{app:data}
In this work we compare our simulation-based predictions with several observational datasets, i.e.,
(i) confirmed MW dwarf satellite galaxies;  
(ii) Ultra-faint compact MW satellite candidates ("UFCSs", systems not yet securely classified as either dwarf galaxies or globular clusters);  
(iii) MW globular clusters; and  
(iv) isolated field dwarf galaxies in the Local Volume.  
Below we summarise the sources, references, and specific parameters adopted for each sample.

The list of confirmed MW dwarf satellites consists of $64$ systems located within $\sim 420$ kpc of the Galactic centre. 
These are
CanisMajor, Columba1, Grus1, Grus2, Horologium1, Horologium2, Hydra1, Indus2, Pegasus3, Phoenix2, Reticulum2, Reticulum3, Sagittarius2, Triangulum2, Tucana3, Tucana4, Antlia2, Aquarius2, Bootes1, Bootes2, Bootes3, Bootes4, Bootes5, CanesVenatici1, CanesVenatici2, Carina, Carina2, Carina3, Centaurus1, Cetus3, ComaBerenices, Crater2, Delve2, Draco, Eridanus2, Eridanus4, Fornax, Hercules, Hydra2, Hydrus1, LMC, Leo1, Leo2, Leo4, Leo5, LeoMinor1, LeoT, Pegasus4, Phoenix, Pictor2, Pisces2, SMC, SagittariusdSph, Sculptor, Segue1, Segue2, Sextans1, Tucana2, UrsaMajor1, UrsaMajor2, UrsaMinor, Virgo1, Virgo2, Willman1.
Of these, only 3 dwarfs are located at distances $>300$ kpc (Eridanus2, LeoT,  Phoenix). We retain them in our comparison to model satellites (defined within $300$ kpc) as they are usually included in the literature, and their properties are entirely typical of the broader confirmed MW satellite population, hence not affecting our conclusions.

The sample of UFCS MW satellite candidates consists of $32$ objects  and follows the compilation of \citet{Smith2024}, whose appendix provides source references for all objects.  These are
\textit{Koposov2, Koposov1, Segue3,} Munoz1, \textit{Balbinot1, Kim1}, Kim2, \textit{Draco2}, Crater, SMASH1, \textit{Kim3}, DESJ0111-1341, DESJ0225+0304, \textit{Eridanus3}, DES1, DES3, PictorI, To1, \textit{Laevens3}, DES4, \textit{PS1-1}, HSC1, DES5, \textit{BLISS1}, Gaia3, \textit{DELVE1}, DELVE2, YMCA-1, DELVE5, \textit{DELVE3, DELVE4}, DELVE6, \textit{UMa3/U1}.

For  confirmed MW dwarf galaxies and UFCS candidates
we use  sky position (RA, Dec), heliocentric distance, absolute $V$-band magnitude $M_V$,  projected half-light radius $R_{\rm half}$, and velocity dispersion measurements $\sigma$
 from the latest update of  \citet{McConnachie2012}'s Nearby Dwarf Galaxy Database\footnote{\url{https://www.cadc-ccda.hia-iha.nrc-cnrc.gc.ca/en/community/nearby/}} (and references therein),  with values updated as in \citet{Cerny2026}. 
Note that we include new $\sigma$ data constraints for a subset of $15$ UFCS systems (those listed in italics above) recently reported  by \citet{Cerny2026}.  Of these, $5$ systems (Balbinot1, DELVE4, Kim1, PS1-1, Koposov2) present resolved or marginally-resolved measurements, while the rest are upper limits.
The $\sigma$ for Tucana3 is an upper limit taken from \citet{Simon2017}.

Stellar masses are computed from $M_V$ assuming a constant stellar mass-to-light ratio of $M_*/L_V = 2$, appropriate for low-mass spheroidal dwarf galaxies \citep{Woo2008} and assuming a solar $V$-band absolute magnitude of $M_{V,\odot}=4.83$ \citep{Allen1973}.
3D half-mass radii $r_{\rm half}$ are estimated  from observed 2D projected half-light radii $R_{\rm half}$ by multiplying the latter by a factor of $4/3$ to correct for spherical projection effects.
Central mean densities are derived from the observed velocity dispersions using the dynamical mass estimator of \citet{Wolf2010} (see Eq.~\ref{eq:dens}).
Galactocentric positions are calculated assuming a Solar position of $R_\odot = 8.29$ kpc.  
For the 47 confirmed MW satellites with available  line-of-sight  velocity and proper-motion measurements, we compute $z=0$ orbital pericentre radii using the MW potential of \citet{Cautun2020}.  Details of the orbital integration procedure are given in sec.~2.6 of \citetalias{Santos-Santos2025}, and the  derived orbital parameters are listed in table~A1 of \citetalias{Santos-Santos2025}.  

For MW GCs we use $M_V$, $R_{\rm half}$, heliocentric distance, and $\sigma$ from the catalogueue of \citet{Harris2010}\footnote{\url{https://physics.mcmaster.ca/~harris/mwgc.dat}}.  
Galactocentric distances are calculated assuming $R_\odot  = 8.0$ kpc.  
We restrict our analysis to clusters with $R_{\rm half} < 1000$ pc, which yields a sample of $157$ objects.  
Pericentre radii are taken from the dynamical models of \citet{Baumgardt2019}\footnote{\url{https://people.smp.uq.edu.au/HolgerBaumgardt/globular/}}, where the orbital integrations were done in the \citet{Irrgang2013} Galactic model. 
In Fig.~\ref{fig:MWrho}, we include only those GCs with both measured velocity dispersions and available orbital parameters. 
We acknowledge that these  pericentric distances for GCs have been derived assuming a   MW potential different to that used for MW satellites shown in the same figure.  We retain these literature values for consistency and  simplicity. Although this mismatch may introduce  a small quantitative uncertainty in the comparison of datapoints,  it has a negligible effect on our interpretation and conclusions.

Finally,  Fig.~\ref{fig:MstRh} includes isolated dwarf galaxies with $M_* < 3 \times 10^{8}\,{\rm M_\odot}$. These are sourced from 
(i) the SPARC (Spitzer Photometry \& Accurate Rotation Curves) database of nearby galaxies \citep{Lelli2016b},  and
(ii) the Local Volume database of \citet{Pace2025}\footnote{\url{https://github.com/apace7/local_volume_database}}, specifically the ``Local Group Field'' and ``Local Group Field Distant'' catalogues.
The SPARC dataset provides homogeneous $K$-band 3.6\,$\mu$m photometry for 175 galaxies.   
We estimate stellar masses assuming a constant stellar mass-to-light ratio of $\Upsilon_{3.6} = 0.5\,{\rm M_\odot/L_\odot}$, following \citet{Lelli2016a}.  Absolute $V$-band magnitudes are derived from 3.6\,$\mu$m luminosities adopting $M_{3.6,\odot} = 3.26$ \citep{Willmer2018} and a colour term of $(V - [3.6]) = 2.5$, consistent with typical values measured for low-surface-brightness galaxies \citep{Schombert2014}.  The  \citet{Pace2025} database includes $M_V$ which we convert to stellar masses assuming a fixed $M_*/L_V=1$ for simplicity.

\section{Evolution of satellite 3D half-mass radii and central densities}\label{app:dens}
\begin{figure*}
\centering
\includegraphics[width=\linewidth]{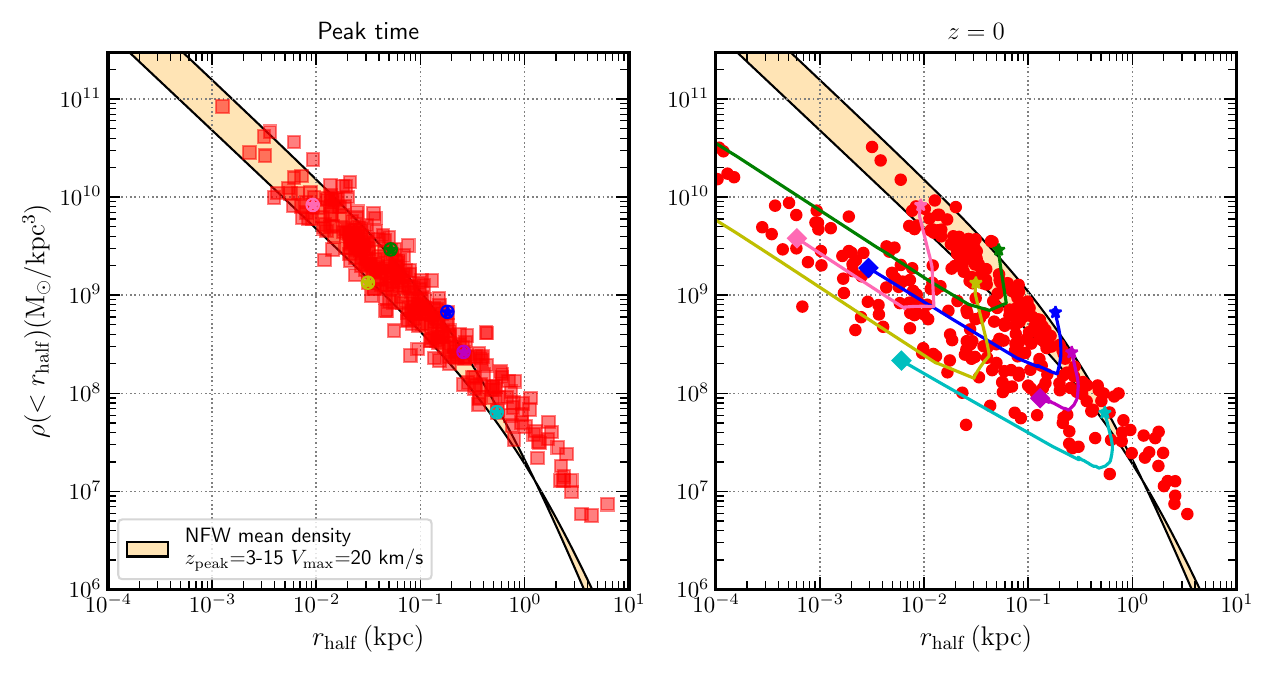}
\caption{
Mean densities within the 3D  half-mass radius $r_{\rm half}$ of Aq-A-L1 satellites at their peak time (left) and at $z=0$ after stripping (right).
The shaded region in the background shows NFW density profiles with $V_{\rm max}=20$ km/s at redshifts $z=3$ and $z=15$, encompassing the range of $z_{\rm peak}$  for Aq-A-L1 satellites. This $V_{\rm max}$ represents the typical peak circular velocity of most satellites (see Fig.\ref{fig:Mstvpk}). 
In the right panel,  $z=0$ results are computed from our stripping model, assuming the "Fiducial" initial luminosity-size relation and the Disc host potential.
Colored lines trace the temporal evolution of selected satellites, with star symbols marking their initial states and diamonds their final $z=0$ positions.
Note that $r_{\rm half}$ also evolves over time (see Fig.~\ref{fig:Lrhalfevo}).
}
\label{figapp:rhorhalf}
\end{figure*}

 Figure~\ref{figapp:rhorhalf} describes the evolution of the central densities of satellites.  The left panel shows, with red square symbols, the satellite mean enclosed densities within the 3D half-mass radius ($r_{\rm half}$) at their peak time.  The background lines correspond to NFW density profiles with $V_{\rm max}\sim 20$ km/s, assuming mean concentrations corresponding to redshifts $z_{\rm peak}=3$ and $15$. This range  roughly encompasses the  peak redshifts of Aq-A-L1 satellites (see Sec.\ref{subsubsec:massloss}).
 At peak time, satellites exhibit densities and half-mass radii that align well within these reference lines,   indicating that the majority of satellite galaxies  are  approximately consistent with  having formed within halos of similar peak mass characterized by $V_{\rm peak}\sim 20$ km/s.  This is in agreement with results shown in Fig.\ref{fig:Mstvpk}.
 
 The right-hand panel shows, with red circles,  the corresponding central densities at $z=0$ after applying the stellar stripping model (we use the "Fiducial" luminosity-size prescription assuming a "Disc" host potential).  Some datapoints stay within the NFW $V_{\rm max}=20$  km/s boundaries while others scatter below, towards lower densities and  smaller $r_{\rm half}$ values.
 
 The evolution of the central density for five example galaxies is overplotted with coloured lines.  Their initial positions are shown with star symbols and  marked in both left and right panels. The end points ($z=0$) are shown as diamonds in the right panel.
 A characteristic evolutionary pattern arises: first, the half-mass radius undergoes a “puffing up” phase as dark matter is stripped from the outskirts (see also Fig.~\ref{fig:Lrhalfevo}),  leading to a decrease in central density.
 Later on,  once mass loss reaches the edge of the galaxy (i.e., when $r_{\rm max} \approx r_{\rm half}$), the stellar component begins to shrink, causing the enclosed central density to rapidly increase.

\bsp	
\label{lastpage}
\end{document}